\documentclass[twoside, a4paper, notoc, 10pt ]{JHEP3}
\usepackage{amssymb}
\usepackage{amsmath}
\usepackage{amsfonts}
\usepackage{epsfig}
\usepackage{bm}
\usepackage{comment}
\usepackage{graphicx}
\usepackage{amsxtra}
\usepackage{cancel}

\newcommand{\mc}{\mathcal}
\newcommand{\bea}{\begin{eqnarray}}
\newcommand{\eea}{\end{eqnarray}}
\newcommand{\vo}{{\cal V}}

\newcommand{\C}[1]{\mathcal{#1}}

\def\beq{\begin{equation}}
\def\eeq{\end{equation}}

\def\arr2[#1,#2][#3,#4]{\left( \begin{array}{cc} #1 & #2 \\ #3 & #4 \end{array} \right)}
\def\vect2[#1,#2]{\left( \begin{array}{c} #1 \\ #2 \end{array} \right)}
\def\V{\mathcal{V}}
\def\2b2[#1,#2][#3,#4]{\left(\begin{array}{cc}#1 & #2 \\ #3 & #4 \end{array} \right)}

\def\beq{\begin{equation}}
\def\eeq{\end{equation}}

\title{Testing String Vacua in the Lab: From a Hidden CMB to Dark Forces in Flux Compactifications}

\author{
Michele Cicoli$\,{}^1$, Mark Goodsell$\,{}^1$, Joerg Jaeckel$\,{}^2$, Andreas Ringwald$\,{}^1$
\\ $^1$Theory Group, Deutsches Elektronen-Synchrotron DESY, Notkestrasse 85, D-22607 Hamburg, Germany \\
$^2$Institute for Particle Physics Phenomenology, Durham University, Durham DH1 3LE, United Kingdom.}

\abstract{We perform a detailed analysis of the phenomenological properties of hidden Abelian gauge bosons
with a kinetic mixing with the ordinary photon within type IIB flux compactifications.
We study the interplay between moduli stabilisation and the Green-Schwarz mechanism that gives mass
to the hidden photon paying particular attention to the r\^{o}le of $D$-terms.
We present two generic classes of explicit Calabi-Yau examples with an isotropic and an anisotropic shape
of the extra dimensions showing how the last case turns out to be very promising to make contact with
current experiments. In fact, anisotropic compactifications lead naturally to a GeV-scale hidden photon (``dark forces'' that can be searched for in beam dump experiments) for an intermediate
string scale; or even to an meV-scale hidden photon (which could lead to a ``hidden CMB'' and can be tested by light-shining-through-a-wall experiments) in the  case of TeV-scale strings. \\

{$e$-mail: \email{michele.cicoli@desy.de}; \email{mark.goodsell@desy.de}; \email{joerg.jaeckel@durham.ac.uk}; \\
\email{andreas.ringwald@desy.de} }}

\preprint{DESY 11-042; IPPP/11/13; DCTP/11/26}
\begin{document}

\tableofcontents

\bigskip

\section{Introduction}
\label{Motiv}

Recently there has been quite some interest in the possibility that there
exist hidden sector particles with masses below a TeV but very weak couplings to Standard Model matter.
They are a common feature of extensions of the standard
model based on supergravity or superstrings.
Extra $U(1)$ gauge
bosons,
so-called hidden photons are a prime candidate for such particles.
At low energies, their interactions with the visible sector
occur primarily via kinetic mixing~\cite{Okun:1982xi,Holdom:1985ag} (studied in string theory in \cite{Dienes:1996zr,Lukas:1999nh,Abel:2003ue,Blumenhagen:2005ga,Abel:2006qt,Abel:2008ai,Goodsell:2009pi,Goodsell:2009xc,Goodsell:2010ie,Heckman:2010fh,Bullimore:2010aj}) with the electromagnetic\footnote{Of course, the mixing is originally with the hypercharge $U(1)$ but after electroweak symmetry breaking this mixing is inherited by the electromagnetic $U(1)$.}\footnote{A massive hidden photon behaves very similar to a $Z^{\prime}$. However, a hidden photon is more specific in the sense that its coupling to Standard Model particles is proportional the ordinary electric charge 
of a particle and the kinetic mixing parameter $\chi$ which is usually small. As the coupling to Standard Model matter is naturally quite small
hidden photons can often be quite light without being in conflict with existing experiments or observations.} $U(1)$,
\begin{eqnarray}
\mathcal{L} \supset  -\frac{1}{4} F^{({\rm vis})}_{\mu \nu} F^{\mu \nu}_{({\rm vis})}
- \frac{1}{4} F^{({\rm hid})}_{\mu \nu} F^{\mu \nu}_{({\rm hid})}+  \frac{\chi}{2} F_{\mu \nu}^{({\rm vis})} F^{({\rm hid}) \mu \nu}
+ m_{\gamma^\prime}^2 A^{({\rm hid})}_{\mu} A^{({\rm hid})\mu}
+A^{({\rm vis})}_{\mu}j^{\mu},
\end{eqnarray}
where $\chi$ is the kinetic mixing parameter and $m_{\gamma^\prime}$ the mass
of the hidden $U(1)$, which may arise via a hidden Higgs or a St\"uckelberg mechanism.
In addition $j^{\mu}$ is the current caused by charged Standard Model matter such as electrons and protons.

\begin{figure}
\begin{center}
\includegraphics[width=0.95\textwidth]{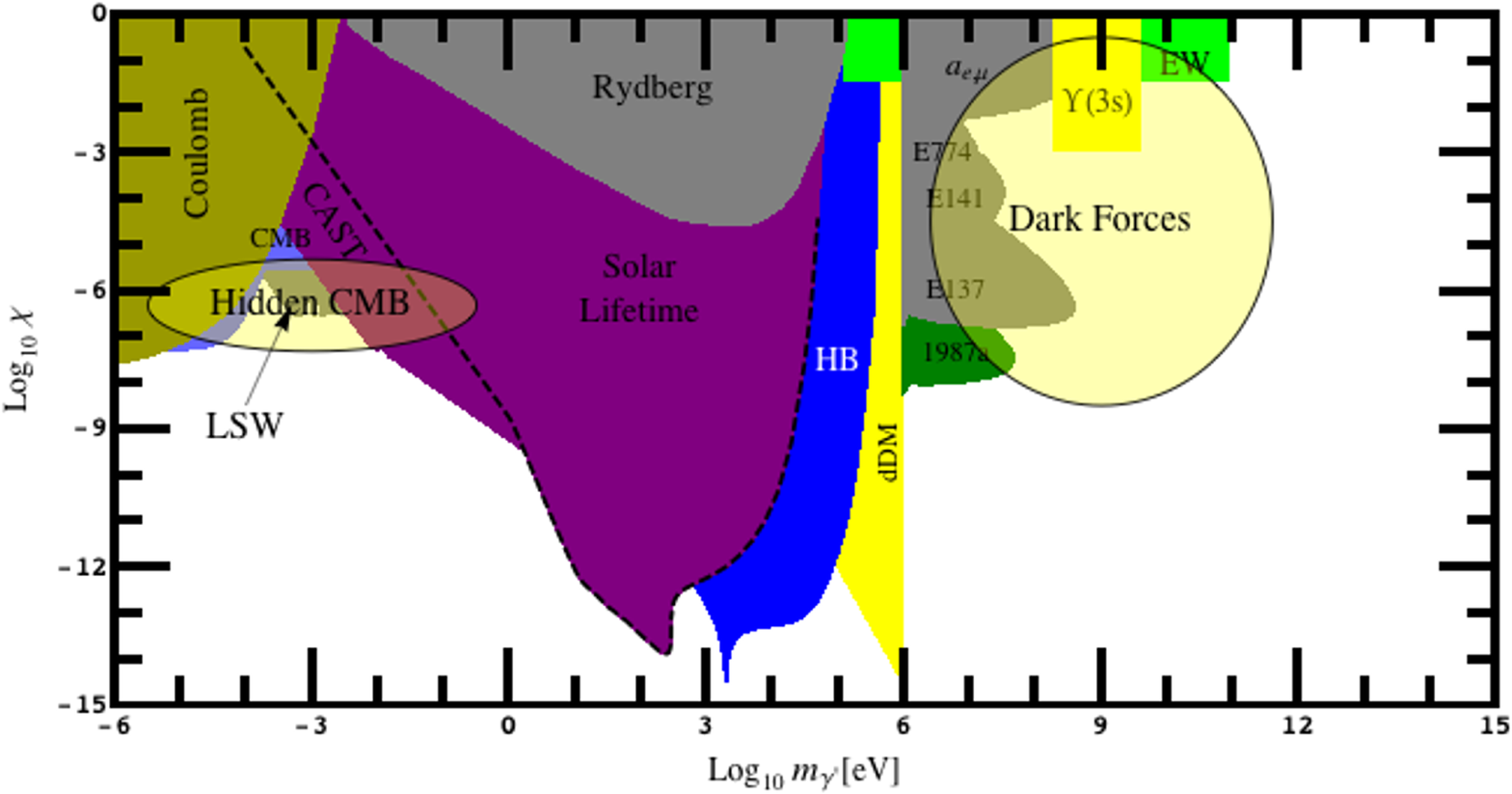}
\end{center}
\vspace{-2.ex}
\caption{Constraints on the kinetic mixing parameter $\chi$ vs. hidden photon mass $m_{\gamma^\prime}$
from astrophysics, cosmology and laboratory experiments.
Phenomenologically interesting regions are marked in yellow. Compilation from~\cite{Jaeckel:2010ni}.}
\label{constraints}
\end{figure}

There are two mass regimes that are of particular phenomenological interest:
the meV range and the GeV range, marked ``hCMB'' and ``Dark Forces'' in Fig.~\ref{constraints}, respectively.
The characteristic behaviour of these two regimes is best understood in slightly different pictures.

At very low masses the most prominent implication of kinetic mixing is that,
similar to neutrino mixing, the propagation and the interaction eigenstates are misaligned.
As a result one expects photon $\leftrightarrow$ hidden photon oscillations~\cite{Okun:1982xi}.
These oscillations could lead to a variety of interesting phenomena.
In the early universe they convert thermal photons into hidden photons, generating a ``hidden CMB'' (hCMB)~\cite{Jaeckel:2008fi}.
Its signature is
an increase in the effective number of relativistic degrees of freedom contributing
to the cosmic radiation density in the era between big bang nucleosynthesis and
recombination beyond the value accounting for the photon and
the three standard neutrino species. Intriguingly, some global cosmological analyses that take into account
precision cosmological data of the cosmic microwave background and of the large scale
structure of the universe appear to require some extra radiation density. The case for this
was strengthened by the recently released WMAP 7 year data whose global analysis points to the
requirement of an equivalent of $\Delta N_\nu^\mathrm{eff}=1.3\pm 0.9$ (68$\%$ C.L.) neutrinos~\cite{Komatsu:2010fb}.

Luckily, hidden photons in the meV range can be nicely searched for in
purely laboratory based laser-light-shining-through-a-wall  experiments~\cite{Ahlers:2007rd,Ahlers:2007qf}, such as
ALPS~\cite{Ehret:2010mh}, BMV~\cite{Fouche:2008jk}, GammeV,
LIPSS~\cite{Afanasev:2008fv}, and OSQAR (cf. the bounds marked ``LSW'' in Fig.~\ref{constraints}),
with great discovery potential in the near future~\cite{Arias:2010bh} and even the possibility of
long distance communication through matter~\cite{Jaeckel:2009wm}.
The discovery potential is also shared by upcoming microwave cavity experiments~\cite{Hoogeveen:1992uk,Jaeckel:2007ch,Caspers:2009cj},
which are currently in the pioneering phase~\cite{Povey:2010hs,Wagner:2010mi}.
In addition, dedicated helioscope searches, e.g. such as SHIPS at the Hamburg Observatory, for hidden photons produced in the sun
could also sensitively explore this region~\cite{Redondo:2008aa}.

At larger masses $\gtrsim$MeV a convenient choice of basis is such that charged Standard Model matter acquires a small
charge under the extra hidden $U(1)$ leading to a ``Dark Forces'' interaction. This type of interactions
can be used to explain a variety of puzzling observations connected to dark matter and astrophysics, such as
the results of DAMA, CoGeNT and PAMELA \cite{ArkaniHamed:2008qn,ArkaniHamed:2008qp,Pospelov:2008jd,Morrissey:2009ur,Feldman:2010wy,Mambrini:2010dq}. Moreover, they provide an interesting explanation
for the deviation $(g-2)_{\mu}$ from the Standard Model prediction~\cite{Pospelov:2008zw}.
These higher masses, too can be sensitively probed in laboratory experiments.
A tool of choice are fixed target experiments~\cite{Bjorken:2009mm} where a high current beam of electrons or protons impacts on
a block of material. A significant number of such experiments are in planning or in trial phases:~DESY (HIPS~\cite{Andreas:2010tp}), MAMI~\cite{Merkel:2011ze} and Jefferson Lab (APEX~\cite{Essig:2010xa}, HPS~\cite{Maruyama:2010} and DarkLight~\cite{Fisher:2010}).

Given this great phenomenological interest and the huge discovery potential for hidden photons,
it is timely to ask whether there are interesting classes of string compactifications which
will lead to kinetic mixing and masses in the ranges described above.

We shall argue that this is indeed the case in a variety of string models
based on type IIB flux compactifications on Calabi-Yau orientifolds with
$D3$/$D7$-branes and $O3$/$O7$-planes \cite{Abel:2008ai,Goodsell:2009xc}, unlike for example the heterotic case \cite{Dienes:1996zr,Goodsell:2010ie} where there is no natural way to suppress the mixing and masses.
The hidden photon can be realised
as an excitation of a space-time filling $D3$ or a $D7$-brane
wrapping an even 4-cycle in the extra dimension separated
from the locus of the $D$-brane hosting the hypercharge $U(1)$
by distances greater than a string length.
This ensures that there are no light states with masses $\lesssim M_s$
charged under both the Standard Model and the hidden gauge groups,
ensuring that the extra $U(1)$ is indeed ``hidden''.

In the $D3$-brane case the kinetic mixing cannot be much smaller than $\chi\simeq 10^{-3}$,
while if a $D7$-brane wraps a large 4-cycle $\tau_{hid}$,
giving rise to a tiny gauge coupling $g^{-2}\simeq \tau_{hid}$, the physical mixing
parameter can be significantly suppressed. Therefore we shall focus on hidden photons
living on these ``milliweak'' or ``hyperweak'' $D7$-branes.
The hidden photon becomes massive
via the Green-Schwarz mechanism by turning on a non-zero world-volume flux.
Some K\"ahler moduli get charged under the hidden $U(1)$ and a particular combination of
the corresponding axions get eaten up by the hidden photon. We stress that this is a truly
stringy mechanism that leads to robust predictions unlike the case of a Higgs mechanism
which depends more heavily on the details of the particular brane construction
\footnote{Moreover, if we aim to generate masses of $\C{O} ({\rm meV})$ with a sizeable mixing $\chi \sim 10^{-6}$,
the very stringent constraints on millicharged particles would require an extremely steep hidden Higgs potential
in order that their masses would be above a MeV.}.

The turning on of gauge fluxes also generates a moduli-dependent Fayet-Iliopoulos term
whose effect on moduli stabilisation has to be taken into account. One of the most
promising ways to fix the geometric moduli in a controlled manner
is given by the type IIB LARGE Volume Scenario \cite{Balasubramanian:2005zx, Cicoli:2008va}.
We shall embed our models into this moduli stabilisation framework since
it does not require fine-tuning of the background fluxes
and can generate exponentially large extra dimensions with the
subsequent possibility to lower the string scale, $\chi$ and $m_{\gamma^\prime}$.

Ref. \cite{Goodsell:2009xc} studied the properties of hidden photons
within the original formulation of the LARGE Volume Scenario
where the compactification is isotropic
in that the largest two-cycle $t_{big}$ is of the order of the cube root of the volume:
$t_{big} \simeq \vo^{1/3}$, but without analysing the r\^{o}le played by $D$-terms in
moduli stabilisation. In this paper we show that $D$-terms for the hyperweak brane
are in general dangerous since they give rise to a run-away for the volume mode.
We propose then a solution where $D$-terms do not cause any decompactification but
dynamically reduce more complicated topologies to the ones
studied in \cite{Goodsell:2009xc}.

The isotropic case leads to nice predictions which however fail to reach the interesting regions of
parameter space corresponding to either hidden CMB or dark forces.
However, once we consider more involved compactification manifolds with a fibration structure as in \cite{Cicoli:2008va},
the desired masses and mixings can be easily accommodated due to the anisotropic shape of the extra dimensions.
In fact, we show that the moduli can be fixed with a single two-cycle that scales as entire volume:
$t_{big}\simeq \vo$. This is complementary to some work in progress \cite{ADDfromStrings},
which shows how maximally anisotropic extra dimensions can be stabilised.
In this case, closed strings propagating along this cycle would be extremely diluted,
and would allow vastly smaller masses for $U(1)$ fields that they couple to.
In this way, the relationship between kinetic mixing and $U(1)$ masses changes dramatically
giving rise to a very interesting phenomenology in the case of ``milliweak'' $D7$-branes
where we find two scenarios:
\begin{itemize}
\item ``Dark force scenario'': the K\"ahler moduli are stabilised
without fine-tuning leading to hidden photons with $m_{\gamma^\prime}\simeq 1$ GeV and $\chi\simeq 10^{-6}$.
The string scale turns out to be intermediate and the Calabi-Yau geometry is
slightly anisotropic;

\item ``Hidden CMB scenario'': fine-tuning the underlying parameters,
the stabilisation of the K\"ahler moduli leads to hidden photons with $m_{\gamma^\prime}\simeq 1$ meV and $\chi\simeq 10^{-6}$.
This corresponds to the extreme case of TeV-scale strings and very anisotropic compactification manifolds.
Furthermore, Kaluza-Klein replicas of the hidden
gauge bosons turn out to be in the dark force mass regime.
\end{itemize}

This paper is organised as follows: section \ref{SECTION:BACKGROUND} provides the essential background information on the properties of D-brane $U(1)$s in IIB compactifications, and establishes our notations; additional description and derivations are presented in the appendix \ref{APPENDIX:Background}. In section \ref{SECTION:EXAMPLES} we describe how anisotropy in IIB compactifications can lead to interesting phenomenology. In section \ref{SECTION:STABILITY} we show how to stabilise the moduli. Section \ref{SECTION:IMPLICATIONS} presents our main results; the predictions for scenarios with stable moduli. It is relatively self-contained so a reader interested only in the predictions testable in experiments can skip the interim sections. Finally we conclude in section \ref{SECTION:CONCLUSIONS}.

\section{Abelian gauge bosons in IIB compactifications}
\label{SECTION:BACKGROUND}

In this section we shall summarise the formulae pertaining to Abelian gauge bosons on D-branes in IIB compactifications that we shall require later. While the material here is not new, we hope that the novel presentation will facilitate their use in model building, specifically for models with stabilised moduli. We present additional explanations and derivations for readers unfamiliar with the material in appendix \ref{APPENDIX:Background}.

Such models are compactified on a Calabi-Yau threefold $\mathcal{M}_6$ which supports a basis of (1,1)-forms $\hat{D}_i$, with K\"ahler form expanded in terms of these forms $J = t^i \hat{D}_i$, intersection numbers
\beq
k_{ijk}=  \int_{\mc{M}_6}\hat{D}_i\wedge\hat{D}_j\wedge\hat{D}_k,
\eeq
and thus the volume of the manifold is $\V = \frac{1}{6} \int J \wedge J \wedge J = \frac{1}{6} k_{ijk} t^i t^j t^k$. Gauge and matter fields are supported on D7 branes wrapping divisors (four-cycles) on the compact space. There is a canonical basis of four-cycles where the Poincar\'e dual two-form is a $\{\hat{D}_i\}$; these have volume $\tau_i = \frac{1}{2} k_{ijk} t^j t^k$ where (neglecting cycles odd under the orientifold) the $\tau_i$ are the real part of the good K\"ahler coordinates for the field theory. A stack of $N$ branes on such a cycle supports a $U(N)$ gauge theory if it is not pointwise invariant under the orientifold projection; if it is then the gauge group is either $Sp(N)$ or $SO(N)$ (depending upon whether the orientifold plane wrapped by the brane is of $O^+$ or $O^-$ type). For a $U(1) \subset U(N)$ the gauge coupling on a brane wrapping such a cycle is given by
\begin{equation}
\frac{2\pi}{g_i^2} = \tau_i.
\label{gaugemaster}\end{equation}
Importantly if the volume of the four cycle is large the gauge coupling can be weak or even hyperweak~\cite{Burgess:2008ri}.

The theory has a classical K\"ahler potential given by
\beq
K_0 \equiv -2 \log \V,
\eeq
which will be modified in the later sections to include corrections due to finite string length and coupling. The above may also be augmented by contributions from collapsed cycles of volume $\tau_0 \approx 0$ of the form $\Delta K = c \frac{(\tau_0)^2}{\V}$. However, derived from the above is the classical K\"ahler metric for the moduli, given by derivatives with respect to the K\"ahler coordinates
\beq
(\C{K}_0)_{ij} \equiv \frac{\partial^2}{\partial \tau_i \partial \tau_j} K_0.
\eeq

\subsection{St\"uckelberg masses}

We shall consider only $U(1)$s supported on D7 branes (rather than R-R $U(1)$s \cite{Arvanitaki:2009hb,Goodsell:2010ie}) since these are generically present in the theory, have the possibility of mixing kinetically with the hypercharge (which must itself be supported on a D7 brane in such models) and crucially may obtain St\"uckelberg masses. The St\"uckelberg mass matrix for the $U(1)$s $a,b$ that do not couple to any cycles odd under the orientifold are given by \cite{Buican:2006sn,Conlon:2008wa,Plauschinn:2008yd,Goodsell:2009xc} \footnote{Allowing for cycles odd under the orientifold plane leads to larger masses and thus less interesting phenomenology; nevertheless the full expression is given in appendix equation (\ref{StmassTotal})}
\begin{align}
\label{Stmass}
m_{ab}^2=& g_a g_b\,  \frac{M_{P}^2}{4\pi^2}  q_{a \alpha} (\C{K}_0)_{\alpha \beta}  q_{b \beta},
\end{align}
where $M_P = 2.4 \times 10^{18}$ GeV is the reduced Planck mass, and where we have defined $q_{ij}$
\beq
q_{ij}= \int_{D_i} \hat{D}_j \wedge \frac{F}{2\pi}  = f_i^k k_{ijk}.
\label{U1charge}
\eeq
which correspond to the ``charges'' of the R-R four-forms under the $U(1)$ supported on brane $i$ with two-form (gauge) flux $\frac{F}{2\pi} = f_i^k \hat{D}_k$ for (half\footnote{The charges can be shifted by a half-integer either in the presence of a discrete $B$-field or in order to cancel Freed-Witten anomalies.}) integers $f_i^k $. Here we are being somewhat cavalier: the above notation somewhat obscures the possibility that a brane may support several $U(1)$ factors.

Crucially the above depends only upon \emph{global} quantities, i.e.  forms and cycles that are defined in the (co)homology of the whole Calabi-Yau, rather than on the branes. In general there will be cycles supported on the branes which are trivial globally, and we should be careful about the correspondence between the global forms and those defining the flux on the branes.

Throughout the text we shall calculate the masses not using the above master formula, but rather using the canonically normalised two-forms to expose where the contributions to the masses come from. As described in the appendix, we define diagonalisation matrices
\beq
(\mc{K}_0^{-1})_{ij} C^j_a =  C_{ia}\lambda_a,\text{ \ \ \ with \ \ \ }(C^t)_a^i C_{ib}= \lambda_a^{-1} \delta_{ab}.
\eeq
which leads to an interaction Lagrangian with canonically normalised fields (\ref{Lfin})
\beq
\mathcal{L}=-\frac{1}{12} \mc{H}_{\mu \nu \rho}^j \mc{H}_j^{\mu \nu \rho}
-\frac{1}{4} \mc{F}_{\mu\nu}^i\mc{F}^{i\,\mu\nu}
+ M_{ij}\mc{D}_2^j \wedge \mc{F}_2^i.
\eeq
As shown in Appendix~\ref{stueckel} this Lagrangian directly leads to St\"uckelberg masses for the U(1) gauge fields given by the sum of contributions from the different canonical forms,
\begin{equation}
m_{ab}^2 = \sum_j M_{aj}M_{bj},
\end{equation}
where
\beq
M_{ij}=\left( g_i \,q_{ip} C^p_j \right) \frac{ M_P}{2\pi} = \left( \frac{q_{ip} C^p_j}{\sqrt{2\pi \tau_i}} \right) M_P.
\label{U1mass}
\eeq

\subsection{Kinetic mixing between hidden and visible photons}

As explained in~\cite{Benakli:2009mk,Goodsell:2009xc}, supersymmetric kinetic mixing is determined by a holomorphic function of only the complex structure moduli which is typically of order a loop factor, but the physical parameter must be multiplied by the gauge coupling of each $U(1)$, giving
\begin{equation}
\label{kinmaster}
\chi\sim \frac{g_{\rm vis}g_{\rm hid}}{16\pi^2},
\end{equation}
where $g_{\rm vis}$ and $g_{\rm hid}$ are the visible and hidden sector gauge groups, respectively.

From this we can immediately see that the kinetic mixing is of the order of $\chi\sim 10^{-3}$ unless the hidden gauge coupling is significantly
weaker than the observed visible ones. This, however, occurs naturally if the $D7$ brane on which the hidden U(1) is realised
has a sizable extent in the extra dimensions. If such a brane has volume  $\tau_{\rm hid}$, this gives (using \ref{gaugemaster})
 \beq
\chi \sim \frac{1}{8\pi}\sqrt{\frac{2\alpha}{ \tau_{\rm hid}}} \sim 0.5 \times \frac{10^{-2}}{\sqrt{\tau_{\rm hid}}}.
\eeq
For large $\tau_{\rm hid}$ the gauge coupling becomes hyperweak and the kinetic mixing can be significantly smaller than the naive expectation.

It is also possible that the supersymmetric kinetic mixing vanishes identically. To determine this, we must have a microscopic understanding of it, and this has so far not been possible on general backgrounds. However, for mutually hidden $U(1)$s it can be understood as arising from exchange of Kaluza-Klein modes of the form fields \cite{Abel:2008ai,Goodsell:2009xc}, and thus if both branes intersect some two-cycle then we expect there to be mixing. This is very similar to the generation of loop corrections to the K\"ahler potential (we are excluding the other contribution in that case - winding modes - since we are insisting that the cycles do not intersect). Furthermore, since it is the excitations of the form fields that mediate the mixing rather than the zero modes, as argued in \cite{Goodsell:2009xc} we expect that they are sensitive to even globally trivial fluxes on the branes, so that even if the hypercharge arises from a GUT structure there should still be mixing.

\subsection{Kaluza-Klein modes}

In addition to the hidden gauge bosons, there will inevitably be a tower of Kaluza-Klein excitations of the hidden gauge field. The determination of the spectrum and couplings of these is in general a complicated task; however, the scaling with the K\"ahler moduli is easily determined and allows us to make a reasonable estimate for the masses of the lightest states (which typically scale as $m_{KK}\sim 1/(length\,\, scale\,\, of\,\, extra\,\, dimension)$). To do this we must examine the geometry of the four-cycle supporting them.  For example, if it is of the form $\mathbb{P}^1 \times \mathbb{P}^1$ then clearly there are two sets of KK modes with different characteristic length scales; if it is of the form $\mathbb{P}^2$ (or blown up with globally trivial exceptional cycles) there is just one. The first example corresponds to $\tau_i = \alpha_{ij} t^j \beta_{ik} t^k$, while the second is $\tau_i = (\alpha_{ij} t^j)^2$. In the latter case, clearly we can take $m_{KK} = \frac{2\pi}{l_s \tau^{1/4}} = \frac{\sqrt{\pi}M_P}{\V^{1/2}\tau^{1/4}} $, while in the former we should take  $m_{KK} = \frac{\sqrt{\pi}M_P}{\V(\alpha_{ij}t^j)^{1/2}} $ assuming that $\alpha_{ij}t^j > \beta_{ik}t^k$.

The other property of interest is whether the Kaluza-Klein modes of the hidden $U(1)$ also kinetically mix with the hypercharge. Indeed, from a calculation on a torus, this seems to be impossible due to Kaluza-Klein momentum conservation. However, this is due to the presence of isometries on the torus which are not present for general geometries. In general the Kaluza-Klein modes are ``unstable'' \cite{Braun:2008jp}, implying that they can mix with the zero mode. We can then ask what the size of the mixing is; here the best estimate we can make is that it is the same order of magnitude as the mixing of the zero modes.

\subsection{Fayet-Iliopoulos terms}

In the presence of a gauge flux the gauge coupling constant $g_i$ is modified to $\frac{2\pi}{g_i^2} = \tau_i - h_i (F_2^c)  \text{Re}(S)$,
where $\text{Re}(S)=e^{-\phi}$ and the flux-dependent factor is given by $h_i (F_2^c)= \frac{f^k f^j k_{ijk}}{2} = \frac{f^j q_{ij}}{2}$
where $q_{ij}$ are the flux-dependent $U(1)$ charges of the K\"{a}hler moduli (\ref{U1charge}).
The Fayet-Iliopoulos term can then be written as:
\beq
\frac{\xi_i}{M_P^2}=\frac{1}{4\pi\vo}\int_{D_i}\left(J\wedge \frac{l_s^2}{2\pi}F^c_2\right)
=\frac{1}{4\pi\vo}t^j f^k k_{ijk}= \frac{q_{ij}}{4\pi}\frac{t^j}{\vo}=-\frac{q_{ij}}{4\pi}\frac{\partial K}{\partial \tau_j}.
\label{FI}
\eeq
Including also the presence of unnormalised charged matter fields $\varphi_j$ (open string states)
with corresponding $U(1)$ charges given by $c_{ij}$, the resulting $D$-term potential looks like
(considering the dilaton fixed at its VEV: $e^{\phi}=g_s$):
\beq
V_D = \frac{g_i^2}{2} \left( \sum_j c_{ij} \varphi_j \frac{\partial K}{\partial \varphi_j} -\xi_i\right)^2
=\frac{\pi}{\left(\tau_i-f^j q_{ij}/(2g_s)\right)} \left( \sum_j c_{ij} \varphi_j \frac{\partial K}{\partial \varphi_j}
+\frac{q_{ij}}{4\pi}\frac{\partial K}{\partial \tau_j}\right)^2.
\label{VD}
\eeq

As we will see later in more detail the significance of the FI-terms is that they have a tendency to destabilise the compactification.
In Sect.~\ref{SECTION:STABILITY} we will discuss ways around this problem.

\subsection{Branes at singularities}

Note that the above still applies for $U(1)$s on branes at singularities. Denoting the two-form corresponding to the canonical class as $\hat{D}_{\mathrm{sing}}$ with K\"ahler modulus $t_{\mathrm{sing}}$, if this is a blow-up mode with only self-intersections then it will only appear in the volume form via a term $a t_{\mathrm{sing}}^3$ and the $U(1)$ will have a string-scale mass. If, however, the singularity intersects some other two-cycle $t_i$ via a term $\V \supset -b t_i t_{\mathrm{sing}}^2$\footnote{Note that such an intersection corresponds to the presence of $N=2$ sectors on toroidal orbifolds, and such intersection forms can be found in the blow-ups \cite{Lust:2006zh}.}, then a flux on $t_{\mathrm{sing}}$ will yield a mass
\begin{align}
m_{\mathrm{sing}}^2 = \frac{M_P^2}{4\pi^2 \V}\frac{1}{b t_i}.
\end{align}
This is particularly interesting since, as described above, in this case the $U(1)$ can mix with the hypercharge if the Standard Model brane also intersects $t_i$ (for example if it is also at a singularity with $\V \supset - b' t_i t_{\mathrm{sing}'}^2$) and also because of the potentially large suppression of the mass if $t_i$ is large; for example if we had a very anisotropic compactification we could have $t_i \sim \V$! However, since the branes sit on singular cycles we cannot suppress the hidden gauge coupling, and so the kinetic mixing will always be of the order $10^{-3}$.

Since we cannot suppress the kinetic mixing (without cancelling it or invoking some fine tuning) in this case, we shall not explore it in detail in the subsequent sections. However, the reader should be aware that such $U(1)$s can be embedded into string compactifications with minimal impact upon moduli stabilisation, and could be interesting for the Dark Forces regime.

\section{Explicit Calabi-Yau examples}
\label{SECTION:EXAMPLES}

\subsection{Isotropic compactifications}
\label{IsComp}

Let us start by studying the case of an orientifold of the Calabi-Yau three-fold given by the
degree 18 hyper-surface embedded in the weighted projective space $\mathbb{C}P_{[1,1,1,6,9]}^{4}$.
The relevant Hodge numbers of this manifold are $h_{1,1} = 2$ and $h_{2,1} = 272$ and its defining equation reads:
\beq
z_1^{18}+z_2^{18}+z_3^{18}+ z_4^3+ z_5^2- 18\psi z_1 z_2 z_3 z_4 z_5
- 3\phi z_1^6 z_2^6 z_3^6=0,
\eeq
where $\psi$ and $\phi$ are the only two complex structure moduli left invariant under the mirror map.
The K\"{a}hler form can be expanded as $J=t_1 \hat{D}_1 +t_2\hat{D}_2$ while the only non-vanishing
intersection numbers are $k_{112}=1$, $k_{122}=6$, $k_{222}=36$.
Thus the overall volume looks like:
\beq
\vo=\frac{1}{6}\int_{CY} J\wedge J\wedge J = \frac{1}{6}\left( 3t_1^2 t_2+18 t_1 t_2^2+36 t_2^3\right).
\eeq
The volumes of the divisors $D_1$ and $D_2$ take the form:
\beq
\tau_1=\frac{1}{2}\int_{D_1} J\wedge J = 3 t_2 (t_1+t_2)\text{, \ \ \ \ \ }\tau_2=\frac{\left( t_1+6 t_2\right)^2}{2},
\eeq
and the Calabi-Yau volume can be rewritten in terms of the 4-cycle volumes as:
\beq
\vo=\frac{1}{9\sqrt{2}}\left(\tau_2^{3/2}-(\tau_2-6\tau_1)^{3/2}\right).
\label{vol}
\eeq
The combination of 4-cycles $D_2-6 D_1$ defines another divisor which is topologically a rigid blow-up mode
resolving a point-like singularity. It is therefore useful to perform the following change of basis: $D_s=D_2-6 D_1$, $D_b=D_2$ and expand the K\"ahler form as $J = t_b \hat{D}_b - t_s \hat{D}_s$.
The intersection numbers in the new basis are very simple since only two of them are non-zero: $k_{sss}=k_{bbb}=36$.
The new expression for the overall volume is:
\beq
\vo= 6 (t_b^3 -  t_s^3) =\frac{1}{9\sqrt{2}}\left(\tau_b^{3/2}-\tau_s^{3/2}\right).  \label{volume}
\eeq
The subscripts $b$ and $s$ stay for `big' and `small' respectively  since we shall be interested in the
large volume limit $t_s\gg t_b$ $\Leftrightarrow$ $\tau_b\gg\tau_s$ $\Leftrightarrow$ $\vo\simeq \tau_b^{3/2}/(9\sqrt{2})$.
In this limit, the K\"{a}hler metric and its inverse look like (defining $\epsilon\equiv\sqrt{\tau_s/\tau_b}\ll 1$):
\beq
 \mc{K}_{0}=\frac{3}{2\tau_b^2}\left(
 \begin{array}{cccc}
 \epsilon^{-1} && -3\epsilon  \\
 -3\epsilon && 2
 \end{array}
 \right)
 \text{ \ \ \ and \ \ \ }
 \mc{K}_{0}^{-1}=\frac{2\tau_b^2}{3}\left(
 \begin{array}{cccc}
 \epsilon && 3 \epsilon^2/2  \\
 3\,\epsilon^2/2 && 1/2
 \end{array}
 \right).
\eeq
The leading order behaviour of the normalised eigenvectors of $\mc{K}_{0}^{-1}$ is:
\beq
\vec{v}_1= \sqrt{\frac{3}{2}}\left\{\frac{1}{\tau_b^{3/4} \tau_s^{1/4}},\frac{3 \tau_s^{3/4}}{\tau_b^{7/4}}\right\}
\text{ \ \ \ and \ \ \ }
\vec{v}_2=\frac{\sqrt{3}}{\tau_b}\left\{3\epsilon^2,1\right\},
\eeq
resulting in the following diagonalising matrix:
\beq
 C^i_j=\frac{1}{\tau_b}\sqrt{\frac{3}{2}}\left(
 \begin{array}{cccc}
 \epsilon^{-1/2} && 3\sqrt{2}\epsilon^2 \\
 3\epsilon^{3/2} && \sqrt{2}
 \end{array}
 \right).
\eeq
Moreover, the internal gauge flux can be expanded in a basis of 2-forms as $F_2^c=f_b \hat{D}_b + f_s \hat{D}_s$.
We are now ready to explore the mass spectrum of possible hidden photons living on $D7$-branes wrapped either
around the large divisor $D_b$ or the small 4-cycle $D_s$.

\subsubsection*{D7 wrapping $D_b$}

We start by analysing the case of a $D7$-brane wrapping the large 4-cycle $D_b$.
Hence we have to set $i=b$ in the general expression (\ref{U1mass}) for the mass of the hidden photon.
Due to the particularly simple structure of the intersection numbers, we obtain:
\beq
M_{bb}=  \left( \frac{1}{\sqrt{2\pi \tau_b}} f_b k_{bbb} C^b_b \right) M_P
= \left(54 \sqrt{\frac{3}{5 \pi}} f_b \right)\frac{M_P}{\tau_b^{3/2}}= \left(3 \sqrt{\frac{6}{5 \pi}} f_b\right) \frac{M_P}{\vo},
\label{Mbb}
\eeq
and
\beq
M_{bs}=  \left( \frac{1}{\sqrt{2\pi \tau_b}}  f_b k_{bbb} C^b_s \right) M_P
= \left(54 \sqrt{\frac{6}{\pi}}  f_b \tau_s\right) \frac{M_P}{\tau_b^{5/2}}\sim f_b \tau_s \frac{M_P}{\vo^{5/3}}.
\eeq
Therefore a particular combination of $\mc{D}_2^b$ and $\mc{D}_2^s$ couples to $F_2$,
but given that $M_{bb}\gg M_{bs}$ for large volume $\vo\gg 1$, we realise that:
\beq
\mathcal{L}_{int}= \left(\frac{M_{bb}}{4} \mc{D}_2^b+\frac{M_{bs}}{4} \mc{D}_2^s\right) \wedge \mc{F}_2 \simeq
\frac{M_{bb}}{4} \mc{D}_2^b \wedge \mc{F}_2\,\,\,\,\Rightarrow\,\,\,\,m_{\gamma^\prime}\simeq M_{bb}\simeq \frac{M_P}{\vo} \simeq \frac{M_s}{\vo^{1/2}}.
\eeq
Furthermore the kinetic mixing between the hidden and the visible photon looks like:
\beq
\chi \simeq 0.5 \cdot \frac{10^{-2}}{\sqrt{\tau_b}}.
\eeq
Inverting this relation, we can eliminate $\tau_b$ in the expression (\ref{Mbb}) and obtain a direct
relation between $\chi$ and $m_{\gamma^\prime}$:
\beq
\chi \simeq 2\cdot 10^{-3} f_b^{-1/3} \, \left(\frac{m_{\gamma^\prime}}{ M_P}\right)^{1/3} \simeq   10^{-9} \cdot f_b^{-1/3} \left(\frac{m_{\gamma^\prime}}{1\  \text{GeV}}\right)^{1/3} .
\label{MchiSC}
\eeq
The exact value of $\chi$ depends on $\tau_b$ whose value should in the end be determined dynamically via moduli stabilisation.
However, regardless of this, the key observation is that once $\chi$ is fixed, the relation (\ref{MchiSC}) also
sets the value of $m_{\gamma^\prime}$ as a function of the flux coefficient $f_b$ which has to be an integer.
This makes it somewhat difficult to reach the interesting regions in the $(m_{\gamma^\prime}, \chi)$-parameter space
corresponding to either hidden CMB or dark forces. In fact, setting $\chi\simeq 10^{-7}$ and $f_b\simeq \mc{O}(1)$
in (\ref{MchiSC}), we obtain $m_{\gamma^\prime}\simeq 10^6$ GeV which is definitely too heavy to explain the extra
relativistic degree of freedom in the CMB and very far beyond detectability for dark forces. Increasing $\chi$
the situation gets even worse since also $m_{\gamma^\prime}$ increases. We finally stress the fact that since $f_b$ has
to be an integer, there is even no room for fine-tuning the mass of the hidden photon.

\subsubsection*{D7 wrapping $D_s$}

Let us now turn to study the case of a $D7$-brane wrapping the small blow-up mode $D_s$.
Setting $i=s$ in the general expression (\ref{U1mass}) for the mass of the hidden photon,
we find:
\beq
M_{sb}=  \left( \frac{1}{\sqrt{2\pi\tau_s}} f_s k_{sss} C^s_b \right) M_P
\sim \frac{f_s}{\sqrt{\tau_s}}  \frac{M_P}{\tau_b}
\sim \frac{f_s}{\sqrt{\tau_s}} \frac{M_P}{\vo^{2/3}},
\eeq
and
\beq
M_{ss}=  \left( \frac{1}{\sqrt{2\pi\tau_s}} f_s k_{sss} C^s_s \right) M_P
\sim \frac{f_s}{\tau_s^{3/4}}  \frac{M_P}{\tau_b^{3/4}}
\sim \frac{f_s}{\tau_s^{3/4}} \frac{M_P}{\vo^{1/2}}.
\eeq
We find again that a particular combination of $\mc{D}_2^b$ and $\mc{D}_2^s$ couples to $F_2$,
but given that $M_{ss}\gg M_{sb}$ for $\vo\gg 1$, we end up with:
\beq
\mathcal{L}_{int}= \left(\frac{M_{sb}}{4} \mc{D}_2^b+\frac{M_{ss}}{4} \mc{D}_2^s\right) \wedge \mc{F}_2 \simeq
\frac{M_{ss}}{4} \mc{D}_2^s \wedge \mc{F}_2\,\,\,\,\Rightarrow\,\,\,\,m_{\gamma^\prime}\simeq M_{ss}\simeq \frac{M_P}{\vo^{1/2}}\simeq M_s.
\eeq
We realise that this case is not very interesting for us
since the Green-Schwarz mechanism generates an $\mc{O}(M_s)$-mass for this Abelian gauge boson
which disappears from the low energy effective field theory. This is the typical behaviour of an anomalous $U(1)$.

\subsection{Anisotropic compactifications}
\label{AnisComp}

We shall now turn to study compactification manifolds whose overall volume is not
controlled by just one large 4-cycle but by several 4-cycles. Therefore in this case
the extra dimensions can have in principle a very anisotropic shape which can crucially
modify the properties of hidden photons. In this section we shall assume an anisotropic
shape of the Calabi-Yau, and show that this property allows us to decouple $m_{\gamma^\prime}$ from $\chi$
being able to reach the more interesting regions of our parameter space corresponding to either
hidden CMB or dark forces.
More precisely, we shall show that the relation (\ref{MchiSC}) has to be modified introducing a
new parameter whose value should be fixed dynamically. In the next sections, we will then
describe a moduli stabilisation mechanism that naturally gives rise to these anisotropic
compactification manifolds.

We shall focus now on the Calabi-Yau manifold defined by the degree 12 hyper-surface
embedded in $\mathbb{C}P_{[1,1,2,2,6]}^{4}$. This Calabi-Yau is a K3 fibration and has $(h^{1,1},
h^{2,1})=(2,128)$ with Euler number $\chi =-252$. Including only the complex
structure deformations that survive the mirror map, the
defining equation is:
\beq
z_{1}^{12}+z_{2}^{12}+z_{3}^{6}+z_{4}^{6}+z_{5}^{2}-12\psi
z_{1}z_{2}z_{3}z_{4}z_{5}-2\phi z_{1}^{6}z_{2}^{6}=0.
\eeq
In terms of 2-cycle volumes the overall volume takes the form:
\beq
\mathcal{V}=t_{1}t_{2}^{2}+\frac{2}{3}t_{2}^{3},
\label{Vol11226}
\eeq
which gives the following relations between the 2- and 4-cycle volumes:
\bea
\tau_{1}=t_{2}^{2}, & \qquad &
\tau_{2}=2t_{2}\left(t_{1}+t_{2}\right),
\nonumber \\
t_{2}=\sqrt{\tau _{1}}, & \qquad & t_{1}=\frac{\tau _{2}-2\tau
_{1}}{2\sqrt{\tau _{1}}}, \label{tT11226}
\eea
Hence the overall volume can be written as:
\beq
\mathcal{V}=\frac{1}{2}\sqrt{\tau _{1}}\left( \tau
_{2}-\frac{2}{3}\tau _{1}\right) .  \label{vol11226}
\eeq
In what follows we shall be interested in anisotropic compactifications for which
$t_1\gg t_2$ $\Leftrightarrow$ $\tau_2\gg\tau_1$, and so the previous relations can be simplified to:
\beq
\mathcal{V}\simeq t_{1}t_{2}^{2}=\frac{1}{2}\sqrt{\tau_{1}}\tau_{2}=t_1\tau_1,
\eeq
with $k_{122}=2$ the only non-vanishing intersection number. The 2-cycle and 4-cycle volumes take the form:
\bea
\tau _{1}=t_{2}^{2}, & \qquad &
\tau_{2}=2t_1 t_{2},
\nonumber \\
t_{2}=\sqrt{\tau _{1}}, & \qquad & t_{1}=\frac{\tau _{2}}{2\sqrt{\tau _{1}}}.
\eea
The cycle $\tau_1$ is a ``milliweak'' cycle, being between a ``small'' and ``hyperweak'' cycle, and arises due to the fibration structure.
In the large volume limit described above,
the K\"{a}hler metric and its inverse look like (defining $\epsilon\equiv\sqrt{\tau_s/\tau_b}\ll 1$):
\beq
 \mc{K}_{0}=\left(
 \begin{array}{cccc}
 \tau_1^{-2} && 0  \\
 0 && 2\tau_2^{-2}
 \end{array}
 \right),
\text{ \ \ \ and \ \ \ }
 \mc{K}_{0}^{-1}=\left(
 \begin{array}{cccc}
 \tau_1^2 && 0  \\
 0 && \tau_2^2/2
 \end{array}
 \right).
\eeq
The normalised eigenvectors of $\mc{K}_{0}^{-1}$ are given by:
\beq
\vec{v}_1= \left\{\tau_1^{-1},0\right\}
\text{ \ \ \ and \ \ \ }
\vec{v}_2=\left\{0,\sqrt{2}\,\tau_2^{-1}\right\}
\eeq
resulting in the following diagonalising matrix:
\beq
 C^i_j=\left(
 \begin{array}{cccc}
 \tau_1^{-1} && 0 \\
 0 && \sqrt{2}\,\tau_2^{-1}
 \end{array}
 \right).
\eeq
Moreover, the internal gauge flux can be expanded in a basis of 2-forms as $F_2^c=f_1 \hat{D}_1 + f_2 \hat{D}_2$.
We are now ready to explore the mass spectrum of possible hidden photons living on $D7$-branes wrapped either
around the ``milliweak'' K3 divisor $D_1$ or the large 4-cycle $D_2$.

\subsubsection*{D7 wrapping $D_2$}

We start by analysing the case of a $D7$-brane wrapping the large 4-cycle $D_2$.
Hence we have to set $i=2$ in the general expression (\ref{U1mass}) for the mass of the hidden photon.
Due to the particularly simple structure of the diagonalising matrix, we find:
\beq
M_{21}=  \left( \frac{1}{\sqrt{2\pi\tau_2}} k_{122}f_2 C^1_1\right) M_P
\sim f_2\frac{M_P}{\sqrt{\tau_2}\tau_1}.
\label{M21}
\eeq
and:
\beq
M_{22}=  \left( \frac{1}{\sqrt{2\pi\tau_2}} k_{122} f_1 C^2_2\right) M_P
\sim f_1 \frac{M_P}{\tau_2^{3/2}}.
\label{M22}
\eeq
Therefore the particular combination of $\mc{D}_2^2$ and $\mc{D}_2^1$ that couples to $F_2$, reads:
\beq
\mathcal{L}_{int}= \left(\frac{M_{22}}{4} \mc{D}_2^2+\frac{M_{21}}{4} \mc{D}_2^1\right) \wedge \mc{F}_2,
\eeq
with the corresponding coefficients that depend on the two different flux parameters $f_1$ and $f_2$.
Given that we are free to turn on the magnetic gauge flux on either $t_1$ or $t_2$,
this implies that when $f_1=0$ only $\mc{D}_2^1$ couples to $F_2$, while viceversa when $f_2=0$ then
the only 2-form that couples to the $U(1)$ field strength is $\mc{D}_2^2$. The generic case when both
$f_1\neq 0$ and $f_2\neq 0$ has the same behaviour of the case with $f_1=0$ since
$M_{21}\gg M_{22}$ in the anisotropic limit $\tau_2\gg \tau_1$. Let us study the two different cases separately.

\bigskip\noindent{\em Gauge flux on $t_1$: $f_2=0$}

\bigskip\noindent
If we turn on a gauge flux only on $t_1$ setting $f_2=0$, we find that $\mc{D}_2^1$ does not couple to $F_2$ since $M_{21}=0$.
Then the interaction Lagrangian takes the form:
\beq
\mathcal{L}_{int}= \frac{M_{22}}{4} \,\mc{D}_2^2 \wedge \mc{F}_2\,\,\,\,\Rightarrow\,\,\,\,m_{\gamma^\prime}=M_{22}.
\eeq
Furthermore the kinetic mixing between the hidden and the visible photon looks like:
\beq
\chi \simeq 0.5 \cdot \frac{10^{-2}}{\sqrt{\tau_2}}.
\label{chiK3}
\eeq
Inverting this relation, we can eliminate $\tau_2$ in the expression (\ref{M22}) and obtain a direct
relation between $\chi$ and $m_{\gamma^\prime}$:
\beq
\chi\simeq  5 \cdot 10^{-3} f_1^{-1/3} \left(\frac{m_{\gamma^\prime}}{M_P}\right)^{1/3} \simeq 4 \cdot 10^{-9} f_1^{-1/3} \left(\frac{m_{\gamma^\prime}}{\text{GeV}}\right)^{1/3},
\eeq
which looks like the same expression for the isotropic case (\ref{MchiSC}). Hence this case does not
look very promising for particle phenomenology.

\bigskip\noindent{\em Generic gauge flux: $f_i\neq 0$ $\forall i=1,2$}

\bigskip\noindent
If a generic flux is turned on with both $f_1\neq 0$ and $f_2\neq 0$, the interaction Lagrangian can
be approximated as:
\beq
\mathcal{L}_{int}= \left(\frac{M_{22}}{4} \mc{D}_2^2+\frac{M_{21}}{4} \mc{D}_2^1\right) \wedge \mc{F}_2
\simeq \frac{M_{21}}{4} \mc{D}_2^1 \wedge \mc{F}_2\,\,\,\,\Rightarrow\,\,\,\,m_{\gamma^\prime}\simeq M_{21},
\eeq
since $M_{21}/ M_{22}\simeq \tau_2/\tau_1\gg 1$. Hence the phenomenological implications of this set-up
are the same as the case in which $f_1=0$ and exactly just $\mc{D}_2^1$ couples to $F_2$. The relation
between $m_{\gamma^\prime}$ and $\chi$ now depends on an additional parameter since:
\beq
\chi \simeq  10^{-2}  \,\frac{\tau_1}{f_2}\, \frac{m_{\gamma^\prime}}{M_P} \simeq  5 \cdot 10^{-21} \frac{\tau_1}{f_2}\, \frac{m_{\gamma^\prime}}{\mathrm{GeV}}.
\eeq
Therefore we have managed to decouple the kinetic mixing parameter from the mass of the hidden photon.
However $\chi$ and $\tau_1$ are not completely independent parameters since the validity of the anisotropic limit
$\tau_2\gg\tau_1$, when expressed in terms of $\chi$ using (\ref{chiK3}), sets a lower bound on $m_{\gamma^\prime}$:
\beq
\tau_1\ll\tau_2\,\,\,\,\Leftrightarrow\,\,\,\,\tau_1\ll 25\cdot 10^{-6}\chi^{-2}
\,\,\,\,\Leftrightarrow\,\,\,\,m_{\gamma^\prime}\gg 4\cdot 10^{24} f_2 \chi^3 \,\text{GeV},
\eeq
which brings us back to phenomenologically uninteresting regions of our parameter space.

The intuitive reason why we are not finding any relevant difference with the isotropic case is that we are
considering a $D7$-brane wrapped around the large 4-cycle $D_2$. In this sense we are not probing the anisotropy
of the Calabi-Yau manifold which, on the other hand, plays a crucial r\^{o}le only if we consider hidden photons
living on the small K3 divisor $D_1$. We shall now turn to study this case showing how we can get more interesting results.

\subsubsection*{D7 wrapping the ``milliweak'' cycle $D_1$}

Let us now turn to study the case of a $D7$-brane wrapping the small K3 divisor $D_1$.
Setting $i=1$ in the general expression (\ref{U1mass}) for the mass of the hidden photon,
the simple form of the diagonalising matrix and the intersection numbers forces $M_{11}=0$.
On the other hand, $M_{12}$ is non-zero and looks like:
\beq
M_{12}=  \left( \frac{1}{\sqrt{2\pi\tau_1}} f_2 k_{122} C^2_2 \right) M_P
\sim f_2 \frac{M_P}{\sqrt{\tau_1}\tau_2}.
\label{M12}
\eeq
Thus if we turn on a gauge flux on $t_1$, we do not couple any 2-form to $F_2$.
This result is in agreement with the general statement that an Abelian gauge boson
can become massive if a non-zero gauge flux is supported on a 2-cycle internal to the
4-cycle wrapped by the corresponding $D7$-brane. Nevertheless in our case the 2-cycle $t_1$
is not internal to $\tau_1=t_2^2$.

Hence we need to turn on a gauge flux on $t_2$, i.e. $f_2\neq0$, which couples $\mc{D}_2^2$ to $F_2$:
\beq
\mathcal{L}_{int}= \frac{M_{12}}{4} \,\mc{D}_2^2 \wedge \mc{F}_2
\,\,\,\,\Rightarrow\,\,\,\,m_{\gamma^\prime}= M_{12}\simeq \frac{M_P}{\vo}\simeq \frac{M_s}{\vo^{1/2}}.
\eeq
Moreover the kinetic mixing between the hidden and the visible photon looks like:
\beq
\chi \simeq 0.5 \cdot \frac{10^{-2}}{\sqrt{\tau_1}}.
\label{chiK3new}
\eeq
Inverting this relation, we can eliminate $\tau_1$ in the expression (\ref{M12}) and obtain
a relation between $\chi$ and $m_{\gamma^\prime}$ which again depends on an additional parameter:
\beq
\chi\simeq 10^{-2}  \,\frac{\tau_2}{f_2}\frac{m_{\gamma^\prime}}{M_P}\,  \simeq 5\cdot 10^{-21}\frac{\tau_2}{f_2}\left( \frac{ m_{\gamma^\prime}}{ \text{GeV}}\right) .
\label{MchiRel}
\eeq
Contrary to the previous case of a $D7$-brane wrapped around $D_2$,
now we have really managed to decouple $\chi$ from $m_{\gamma^\prime}$ since the anisotropic limit
$\tau_2\gg\tau_1$, when expressed in terms of $\chi$ using (\ref{chiK3new}), now gives just
an irrelevant upper bound on $m_{\gamma^\prime}$:
\beq
\tau_1\ll\tau_2\,\,\,\,\Leftrightarrow\,\,\,\,\tau_2\gg 25\cdot 10^{-6}\chi^{-2}
\,\,\,\,\Leftrightarrow\,\,\,\,m_{\gamma^\prime}\ll 4 \cdot 10^{24} f_2 \chi^3 \,\text{GeV}.
\eeq
Setting $\chi\sim 10^{-7}$ and $f_2\simeq \mc{O}(1)$, this upper bound becomes $m_{\gamma^\prime}\ll 10^5$ GeV
without ruling out any interesting region of our parameter space. Clearly, increasing $\chi$ this upper bound
becomes even less stringent.

Hence we have found a very promising set-up in an anisotropic compactification which
opens up the possibility to reach regions of the $(m_{\gamma^\prime},\chi)$-parameter space
that are very appealing for hidden CMB and dark forces. However in order
to be able to get a sensible prediction, one needs to understand the dynamics of the
extra dimensions and fix the value of $\tau_1$ and $\tau_2$. In the next section we shall present
a moduli stabilisation mechanism that will allow us to derive a concrete
prediction for $m_{\gamma^\prime}$ and $\chi$ in this set-up in a completely top-down approach from string theory.

\section{Stabilisation of the extra dimensions}
\label{SECTION:STABILITY}

In this section we shall follow \cite{Cicoli:2008va} and present a moduli
stabilisation mechanism that naturally leads to anisotropic
compactifications with both $\tau_1$ and $\tau_2$ fixed at large values
in string units. We shall then explore the phenomenological implications
of this class of string vacua and show that they can give rise to
two interesting scenarios for hidden photons:
\begin{itemize}
\item Considering natural values of the underlying parameters
leads to hidden photons with $m_{\gamma^\prime}\simeq 1$ GeV and $\chi\simeq 10^{-6}$,
for intermediate string scale $M_s\simeq 10^{12}$ GeV. These values of
the kinetic mixing parameter and the mass of the hidden photon
are in the region of parameter space that will be soon tested by the next beam dump and fixed target experiments,
and produce a particle with the right properties to explain the Dark Forces phenomena.
On top of that, an intermediate string scale naturally yields TeV-scale supersymmetry,
a QCD axion with a decay constant $f_a\simeq M_s$ within the allowed window,
and the right Majorana scale for right handed neutrinos.

\item Fine tuning the underlying parameters leads to
hidden photons with $m_{\gamma^\prime}\simeq 1$ meV and $\chi\simeq 10^{-6}$
for the extreme case of a TeV string scale $M_s\simeq 1$ TeV. These values of
the kinetic mixing parameter and the mass of the hidden photon
yield a new particle with the right properties to account for the observational evidence
of an extra relativistic degree of freedom in the CMB.

Moreover, in this case there is no need to have TeV-scale supersymmetry since the
hierarchy problem is solved by the low string scale,
that would also allow to probe string scale physics at the LHC.\footnote{Here the string scale is simply defined as $M_s \equiv 1/l_s$. However, the string resonances occur at multiples of $2\pi/l_s$; the lowest such universal states should be seen therefore at $2\pi M_s$.}
We also find as an accidental bonus in this case that the Kaluza-Klein modes of the hidden gauge boson have masses of the right order of magnitude to be observed in the ``Dark Forces'' regime (assuming they also kinetically mix with the zero modes):
\begin{align}
m_{KK} \sim \frac{M_s}{\tau_1^{1/4}} \sim 1 - 10 \;\mathrm{GeV}.
\end{align}

Large radiative corrections to the moduli masses due to the absence of supersymmetry
and the weakness of some of the moduli couplings (which might be much weaker than $1/M_P$)
due to the geometric separation between different 4-cycles within the Calabi-Yau,
lead to no conflict with fifth force experiments \cite{Burgess:2010sy, ADDfromStrings}.
\end{itemize}

We shall now focus on the stabilisation of all the geometric moduli
which emerge in the low energy effective field theory of type IIB
compactified on an orientifold of the Calabi-Yau three-fold
given by the addition of a blow-up mode to the geometry $\mathbb{C}P^{4}_{[1,1,2,2,6]}[12]$
studied in the previous sections. Explicit compact Calabi-Yau examples
with these features have been recently constructed in \cite{explicit}.
We point out that, as we shall see later on, the inclusion of an extra blow-up
mode is required to guarantee the existence of
controlled large volume solutions. Therefore the Calabi-Yau
volume in terms of its three K\"ahler moduli looks like:
\begin{equation}
 \mathcal{V} = \lambda_{1}t_{1}
 t_{2}^{2} + \lambda_{3}t_{3}^{3}
 = \alpha \left( \sqrt{\tau _{1}}\tau _{2} - \gamma \tau _{3}^{3/2}\right)
 = t_1\tau_1-\alpha\gamma\tau_3^{3/2},  \label{hhh}
\end{equation}
with the constants $\alpha$ and $\gamma$ which depend on the intersection numbers
and are taken to be positive and typically of order unity.
In order to obtain light hidden photons we are interested in large values of the
overall volume, and so we shall work in the parameter regime:
\begin{equation}
\label{V0hier}
 \vo\simeq \alpha\sqrt{\tau_1}
 \; \tau_2 \gg \alpha\gamma\tau_3^{3/2} \gg 1.
\end{equation}
Regarding the relative size of each K\"{a}hler modulus,
we shall consider the limit
$\tau_2 \gg \tau_1 \gg \tau_3$ $\Leftrightarrow$ $t_1 \simeq \tau_2/\sqrt{\tau_1} \gg t_2
\simeq \sqrt{\tau_1} \gg t_3 \simeq \sqrt{\tau_3}$, corresponding to
the interesting anisotropic case having the two dimensions of the base,
spanned by the cycle $t_1$, hierarchically larger than the other
four of the K3 fibre, spanned by $\tau_1$. 

\subsection{$F$-term potential}

\subsubsection{Tree-level effective action}

The geometric moduli of the $\mathcal{N} = 1$ 4D supergravity
obtained as a low-energy limit of type IIB string theory
compactified on a Calabi-Yau orientifold,
include $h_{1,1}$ K\"{a}hler moduli defined in (\ref{defT})
which parameterise the size of the internal manifold,
$h_{2,1}$ complex structure moduli $U_{\alpha}$ which
parameterise the shape of the Calabi-Yau,
and the axio-dilaton $S=e^{-\phi}+i C_{0}$,
defined in terms of the R-R 0-form $C_0$ and the dilaton $\phi$,
whose vacuum expectation value sets the string coupling: $g_s^{-1}= \langle\text{Re}(S)\rangle$.

The tree level K\"{a}hler potential $K_{tree}$ takes the factorised form (setting $M_P=1$ for the time being):
\begin{equation}
K_{tree}(T+ \bar{T},S+\bar{S},U)=-2\ln \mathcal{V}-\ln \left(
S+\bar{S}\right) -\ln \left( -i\int\limits_{CY}\Omega \wedge
\bar{\Omega}\right) ,  \label{Ktree}
\end{equation}
where $\mathcal{V}$ depends only on $(T+ \bar{T})$ and $\Omega $ is
the Calabi-Yau holomorphic $(3,0)$-form which is a function of
the complex structure moduli.

A key ingredient to fix most of the geometric moduli is the turning on
of background fluxes $G_3 = F_3 + i S H_3$, where
$F_3$ and $H_3$ are respectively the R-R and NS-NS 3-form fluxes \cite{Giddings:2001yu}.
These fluxes generate a tree-level superpotential which takes the
Gukov-Vafa-Witten form \cite{Gukov:1999ya}:
\begin{equation}
W_{tree}(S, U)=\int\limits_{CY}G_{3}\wedge \Omega ,
\label{Wtree}
\end{equation}
As $W_{tree}$ is independent of the K\"{a}hler moduli,
the $\mc{N}=1$ $F$-term supergravity scalar potential looks like:
\begin{equation}\label{scalarpotential}
V_F=e^{K}\left[ \sum_{S,U}K^{\alpha\bar{\beta}}D_{\alpha}W D_{\bar{\beta}}\bar{W}
+ \left(\sum_{T} K^{i\bar{j}}K_{i}K_{\bar{j}}-3\right)\left\vert W\right\vert ^2\right].
\end{equation}
Due to the no-scale property of the tree-level K\"{a}hler potential (\ref{Ktree}),
$\sum_{T} K^{i\bar{j}}K_{i}K_{\bar{j}}=3$, all the $T$-moduli are exactly flat directions
at semiclassical level. Thus one is left with a semi-positive definite scalar potential for
the $S$ and $U$-moduli which admits a Minkowski minimum for $D_{S}W=D_{U}W=0$.

If we are then interested in fixing the K\"{a}hler moduli at subleading order
via perturbative and non-perturbative corrections, we can safely set the dilaton and
the complex structure moduli equal to their vacuum expectation values.
Then the superpotential is constant, $W = \langle
W_{tree} \rangle\equiv W_0$ and the K\"{a}hler potential is
$K=K_{cs} - \ln\left({2}/{g_{s}}\right) + K_0$ with:
\beq
K_0 = -2\ln \mathcal{V} \qquad \hbox{and} \qquad e^{-K_{cs}} =
\left\langle -i\int_{CY} \Omega \wedge \bar{\Omega}\right\rangle
\, . \label{K0expr}
\eeq

\subsubsection{Leading order corrections}

Let us now consider the leading order corrections to the tree level action
which lift the remaining flat directions.
These are the leading order $\alpha'$ corrections to $K$ \cite{Becker:2002nn}:
\begin{equation}
  K=K_0+\delta K_{\alpha'}= -2\ln \left( \mathcal{V}+\frac{\xi }{2g_{s}^{3/2}}\right)
  \simeq -2\ln\mathcal{V}-\frac{\xi}{g_s^{3/2}\mathcal{V}},
 \label{Kalpha'}
\end{equation}
and non-perturbative corrections to $W$:
\begin{equation}
 W = W_{0} + A_{3} e^{-a_{3} T_{3}} \,.
 \label{sp}
\end{equation}
The correction (\ref{Kalpha'}) comes from the reduction of the $\mc{O}(\alpha'^3)\mc{R}^4$ 10D term and
corresponds to higher derivative corrections in the effective supergravity description.
The parameter $\xi$ is given by $ \xi =-\frac{\chi \zeta (3)}{2(2\pi
)^{3}}$, where $\chi=2\left(h_{1,1}-h_{2,1}\right)$ is the Calabi-Yau
Euler number, and the Riemann zeta function is $\zeta(3) \simeq
1.2$.

On the other hand, the non-perturbative correction to the superpotential (\ref{sp})
can be generated wrapping the blow-up mode $D_3$ with
either a Euclidean $D3$-brane instanton (in which case $a_3=2\pi$)
or a stack of $D7$-branes supporting an $SU(N)$ gauge theory that undergoes
gaugino condensation (in which case $a_3=2\pi/N$). Notice that the fact that
$D_3$ is a rigid divisor guarantees the existence of such kind of non-perturbative effects.
The coefficient $A_3$ corresponds to threshold effects and it depends on $U$ and $D3$-position moduli,
but not on the K\"{a}hler moduli.

Therefore the $F$-term scalar potential at leading order in a large volume expansion, looks like:
\begin{equation}
 V_F = \frac{g_s e^{K_{cs}}M_P^4}{8\pi}\left[\frac{8 \, a_{3}^{2}A_3^2}{3\alpha\gamma}
 \left( \frac{\sqrt{\tau _{3}}}{\mathcal{V}}
 \right) e^{-2a_{3}\tau_3}
 +4 W_{0}a_{3} A_3 \cos (a_3 b_3) \left( \frac{\tau _{3}}{\mathcal{V}^{2}}
 \right) \, e^{-a_{3}\tau_3}
 +\frac{3 \, \xi W_0^2}{4 g_s^{3/2}\mathcal{V}^{3}} \right],
 \label{Pot}
\end{equation}
where we have explicitly included the right prefactor obtained from dimensional reduction (see appendix of \cite{Burgess:2010bz}).
Taking both $W_0$ and $A_3$ to be real and positive without loss of generality,
the minimum for the axion $b_3$ is at $b_3 = k\pi/a_3$ with $k\in\mathbb{Z}$.
The potential for $\tau_3$ and $\vo$ then admits a minimum for $\xi>0$ (i.e. $h_{2,1}>h_{1,1}=3$)
located at:
\begin{equation}
 \langle \tau _{3}\rangle = \frac{1}{g_s}\left(
 \frac{\xi}{2\, \alpha \gamma } \right) ^{2/3}
 \qquad \hbox{and} \qquad
 \langle
 \mathcal{V}\rangle = \left( \frac{ 3 \,\alpha \gamma }{4a_{3}A_3}
 \right) W_0 \, \sqrt{\langle \tau _{3}\rangle }
 \; e^{a_{3} \langle \tau_3
 \rangle }\text{\ .}  \label{LVSmin}
\end{equation}
This is the typical non-supersymmetric AdS minimum of LARGE volume scenarios \cite{Balasubramanian:2005zx, Cicoli:2008va}.
Supersymmetry is broken spontaneously by non-zero $F$-terms of the K\"{a}hler moduli \cite{Conlon:2006wz} and
the minimum is found without fine-tuning the background fluxes, i.e. setting $W_0\simeq \mc{O}(1)$.
The exponentially large volume allows to explain many
hierarchies observed in nature and guarantees that the low-energy
effective field theory is under good control.

Due to the fact that, within this level of approximation,
$V_F$ depends only on two of the three original K\"{a}hler moduli,
$V_F=V_F(\vo,$\ $\tau_{3})$, we have been able to fix only a particular combination
of $\tau_1$ and $\tau_2$ corresponding to the overall volume.
The potential along the other orthogonal combination is therefore so far completely flat.
This direction played the r\^{o}le of the inflaton in \cite{Cicoli:2008gp,Cicoli:2009zh,Cicoli:2010ha,Cicoli:2010yj}
and can be lifted via subleading contributions to (\ref{Pot})
coming from string-loop corrections to the K\"{a}hler potential.

\subsubsection{Subleading order corrections}
\label{StringLoopCorr}

The next to leading order correction to the flat tree-level potential for
the $T$-moduli comes from 1-loop open string contributions to the K\"{a}hler potential.
Their form has been explicitly computed only in the simple case of
$\mathcal{N}=1$ toroidal orientifolds and looks like \cite{Berg:2005yu}:
\begin{equation}
 \delta K_{(g_{s})}=\delta K_{(g_{s})}^{KK}+\delta K_{(g_{s})}^{W},
\end{equation}
where $\delta K_{(g_{s})}^{KK}$ can be interpreted from the closed string point of view,
as coming from the exchange of Kaluza-Klein modes between $D7$
and $D3$-branes or non-intersecting stacks of $D7$-branes,
while $\delta K_{(g_{s})}^{W}$ is due to the
exchange of winding strings between intersecting stacks of $D7$-branes.
Assuming that all the three 4-cycles of the torus are wrapped by $D7$-branes,
these two corrections read:
\begin{equation}
 \delta K_{(g_{s})}^{KK}= -\frac{1}{128\pi^{4}}
 \sum\limits_{i=1}^{3}
 \frac{\mathcal{E}_{i}^{KK}(U,\bar{U})}{\hbox{Re}\left( S\right)
 \tau _{i}},\text{ \ \ \ and \ \ \ }
 \delta K_{(g_{s})}^{W}=-\frac{1}{128\pi^{4}}
 \sum\limits_{i=1}^{3}\left.
 \frac{\mathcal{E}_{i}^{W}(U,\bar{U})}{\tau _{j}\tau_{k}}
 \right\vert _{j\neq k\neq i},
\end{equation}
where the functions $\mathcal{E}_{i}(U,\bar{U})$
are encoding the very complicated dependence of these
corrections on the complex structure moduli.

These results have been used to conjecture the form
of the string loop corrections to $K$ for an arbitrary
Calabi-Yau compactification using two observations:
the interpretation as the tree-level propagation of Kaluza-Klein
and winding modes respectively, and the
Weyl rescaling needed to convert the
string computation to 4D Einstein frame \cite{Berg:2007wt}.
The final proposal is:
\begin{equation}
 \delta K_{(g_{s})}^{KK}=\sum_i
 \frac{\mathcal{C}_{i}^{KK}(U,\bar{U})m_{KK}^{-2} }{\hbox{Re}
 \left( S\right) \mathcal{V}} = \sum_i
 \frac{\mathcal{C}_{i}^{KK}(U,\bar{U})\left( a_{il}t^{l}\right)
 }{\hbox{Re} \left( S\right) \mathcal{V}},  \label{UUU}
\end{equation}
and:
\begin{equation}
 \delta K_{(g_{s})}^{W}= \sum_i\frac{\mathcal{C}_{i}^{W}(U,\bar{U})m_{W}^{-2}}{\mathcal{V}}
 =  \sum_i\frac{\mathcal{C} _{i}^{W}(U,\bar{U})}{\left(
 a_{il}t^{l}\right) \mathcal{V}}. \label{UUUU}
\end{equation}
The linear combination $\left(a_{il}t^l\right)$ of the volumes of the basis 2-cycles $t_l$,
in (\ref{UUU}) gives the 2-cycle that is transverse to the 4-cycle wrapped by the $i$-th $D7$-brane,
whereas in (\ref{UUUU}) it gives the 2-cycle where the two $D7$-branes
intersect. The unknown functions $\mathcal{C}^{KK}_i (U, \bar{U})$ and
$\mathcal{C}^W_i (U, \bar{U})$ can be simply regarded
as free parameters since the complex structure moduli are already
stabilised at the semi-classical level by background fluxes.

A key property of these corrections is that their leading contribution
to the scalar potential is vanishing, leading to an `extended no-scale structure'
which has a nice low-energy interpretation in terms of the Coleman-Weinberg potential \cite{Cicoli:2007xp}.
This leading order cancellation is crucial to render $\delta V_{(g_s)}$ subdominant
with respect to $\delta V_{(\alpha')}$. In fact, the first non-vanishing contribution to
the scalar potential of the corrections (\ref{UUU}) and (\ref{UUUU}) reads \cite{Cicoli:2007xp}:
\begin{equation}
 \delta V_{\left( g_{s}\right) }^{1-loop}=\left[
 \left( g_s \mathcal{C}_{i}^{KK} \right)^2 K^0_{i\bar\imath} - 2 \delta
 K_{(g_{s})}^{W}\right] \frac{W_{0}^{2}}{\mathcal{V}^{2}}.
 \label{V at 1-loop}
\end{equation}
This contribution is subdominant relative to the leading $\alpha'$ correction,
since it scales as $\delta V_{(g_s)}\sim \vo^{-3} t^{-1}$ while
$\delta V_{(\alpha')}\sim g_s^{-3/2}\vo^{-3}$. Hence their ratio behaves as
$\delta V_{(\alpha')}/\delta V_{(g_s)}\sim g_s^{-3/2}t\gg 1$
since we require $g_s\ll 1$ to be in the perturbative regime and $t\gg 1$ (in string units)
to trust the effective field theory.

\bigskip

We shall now apply these general results to our K3 fibred Calabi-Yau
case where we wrap a stack of $D7$-branes around each divisor. The stack of $D7$-branes
wrapped around $D_3$ is needed to generate the non-perturbative effects via gaugino condensation,
while the two stacks wrapped around $D_1$ and $D_2$ generate
the loop corrections needed to fix the remaining flat direction and provide hidden
$U(1)$ gauge bosons.

\begin{figure}[ht]
\begin{center}
\includegraphics[width=0.85\textwidth]{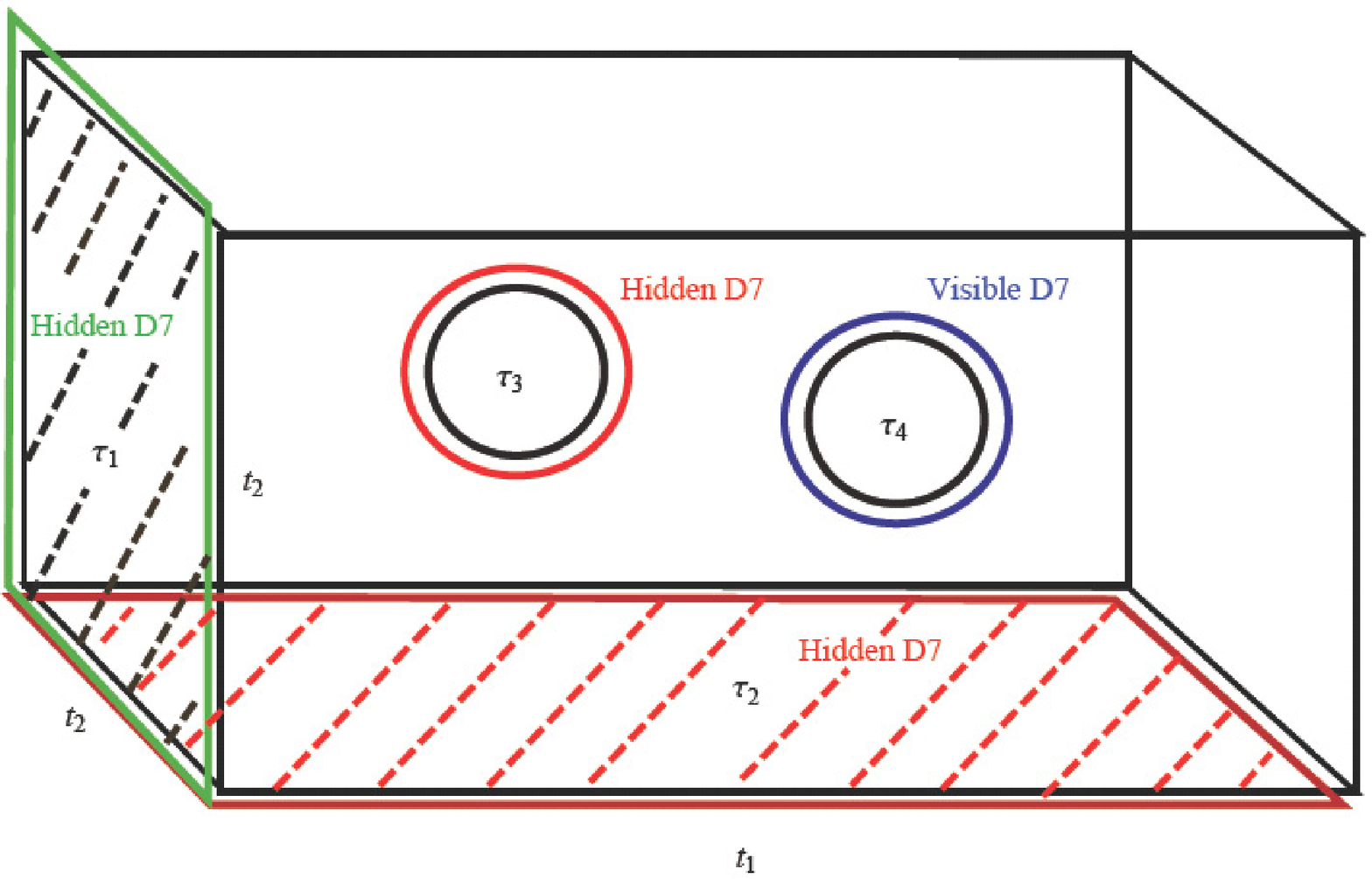} \caption{Pictorial view of the K3 fibred Calabi-Yau three-fold
and the brane set-up under consideration. Four-cycles are shown as surfaces and two-cycles as lines.} \label{Fig4}
\end{center}
\end{figure}

The general formula (\ref{V at 1-loop}) then gives rise to four corrections to the scalar potential (\ref{Pot}):
\begin{equation}
 \delta V_F= \frac{g_s e^{K_{cs}}M_P^4}{8\pi}\left[\delta V^{KK}_{(g_{s}), \tau_{1}}+\delta V^{W}_{(g_{s}),
 \tau_{1}\cap\tau_{2}}+\delta
 V^{KK}_{(g_{s}), \tau_{2}}+\delta V^{KK}_{(g_{s}),\tau _{3}}\right],
 \label{dVF}
\end{equation}
which have the form:
\begin{eqnarray}
 \delta V_{(g_{s}),\tau _{1}}^{KK} &=&
  \frac{A W_{0}^{2}}{\tau_1^2 \mathcal{V}^{2}},\text{ \ \ \ \ }
  \delta V_{(g_{s}),\tau _{1}\cap\tau _{2}}^{W}
 =-  \frac{B W_{0}^{2}}{\mathcal{V}^{3}\sqrt{\tau_1}}, \label{LOOP} \\
 \delta V_{(g_{s}),\tau _{2}}^{KK} &=&
 \frac{C W_{0}^{2} \tau_1}{\mathcal{V}^4},\text{ \ \ \ \ \ }
 \delta V^{KK}_{(g_{s}),\tau _{3}}=
 \frac{D W_0^{2}}
 {\mathcal{V}^{3}\sqrt{\tau_{3}}},  \notag
\end{eqnarray}
with:
\beq
 A = \left(g_{s} C_{1}^{KK}\right) ^{2}>0, \text{ \ \ }
 B = 4\alpha C_{12}^{W},\text{ \ \ }
 C = 2\,\left(\alpha g_{s}C_{2}^{KK}\right)^{2}>0,\text{ \ \ }
 D = \left(g_{s} C_{3}^{KK}\right) ^{2}>0. \notag
\eeq
Notice that the last term in (\ref{dVF}) can be safely neglected since it does not introduce a
dependence on the remaining flat direction, which, on the other hand, is lifted by the first three terms.
In fact, minimising $\delta V_F$ with respect to $\tau_1$ at fixed $\vo$ and $\tau_3$,
we find:
\beq
  \frac{1}{\langle\tau_1\rangle^{3/2}} = \left( \frac{B}{8 A \langle\vo\rangle} \right)
  \left[ 1 + (\hbox{sign} \, B) \sqrt{1 + \frac{32 AC}{B^2}}
  \right] \label{tau1soln1} \,,
\eeq
which, when $32 A C\ll B^2$ $\Leftrightarrow$ $4\, g_s^4 \left(C_1^{KK} C_{12}^W\right)^2\ll \left(C_2^{KK}\right)^2$,
reduces to:
\beq
  \langle \tau_1 \rangle
 \simeq \left(-\frac{B \langle\vo\rangle}{2C} \right)^{2/3}
  \quad \hbox{if $B<0$}
  \qquad \hbox{or} \qquad
 \langle \tau_1 \rangle \simeq
 \left(\frac{4A \langle\vo\rangle}{B} \right)^{2/3}
  \quad \hbox{if $B>0$}. \label{tau1soln2}
\eeq
Choosing for definiteness $B>0$, we can reexpress the relation (\ref{tau1soln2})
in terms of $\tau_1$ and $\tau_2$ as:
\beq
\langle \tau_1 \rangle =  \kappa \,\langle\tau_2\rangle,
\text{ \ \ \ \ with \ \ \ \ }\kappa\equiv \frac{\left(g_s C_1^{KK}\right)^2}{C_{12}^W}.
\eeq
We finally point out that due to the incompatibility between chirality and non-perturbative 
effects \cite{Blumenhagen:2007sm}, the visible sector (the Standard Model or any generalisation thereof) 
cannot be wrapped around $\tau_3$ but 
it has to be supported by another 
blow-up mode which we shall call $\tau_4$ (see Fig.~\ref{Fig4}). This additional 4-cycle can be fixed either via 
$D$-terms \cite{Blumenhagen:2007sm} or via string loop corrections \cite{Cicoli:2008va}.

\subsection{$D$-term potential}

As we have seen in the previous sections, every time a gauge flux is turned on to
give a St\"{u}ckelberg mass to the hidden photon, also a moduli dependent Fayet-Iliopoulos term
gets generated. In this case, it is therefore inconsistent to study moduli stabilisation focusing just on the
$F$-term scalar potential and neglecting the $D$-term contribution.

When $D$-terms are properly taken into account, they turn out to dominate over $V_F$ generically
giving rise to a dangerous run-away behaviour for the overall volume mode.
Here are some possible way-outs:
\begin{itemize}
\item In the absence of matter fields charged under the $U(1)$, there are two solutions:

\begin{enumerate}
\item Fine-tune the coefficients of $V_D$. Given that all these are $\mc{O}(1)$ numbers, the
only way to achieve small $D$-terms is via warping. Then $V_D$ could be used to turn the AdS minimum
into a Minkowski or slightly de Sitter one. However this solution is not very satisfactory
since it is hard to envisage a situation
where a $D7$-brane wrapping a large 4-cycle is in a highly warped region. In addition one should
check if also $V_F$ is affected by warping. Finally it might still be complicated to estimate the
new predictions for $m_{\gamma^\prime}$ and $\chi$ since the kinetic terms of the 2-forms which couple to $F_2$,
cannot be explicitly canonically normalised in the presence of warping.

\item Consider more complicated topologies with intersecting large 4-cycles.
In this case, requiring a vanishing Fayet-Iliopoulos term fixes
a particular combination of the large divisors and the contribution of $V_D$ gets cancelled dynamically.
One should check that indeed no matter gets generated at the intersection of the two 4-cycles.
\end{enumerate}

\item In the presence of matter fields charged under the $U(1)$, there are two situations:

\begin{enumerate}
\item If all the $U(1)$-charges of the matter fields have the same sign, the $U(1)$ is anomalous and
each scalar acquires a vanishing vacuum expectation value. Hence, as far moduli stabilisation is concerned,
we get back to the situation above where we did not consider any matter field, but with the additional
phenomenological constraint of avoiding the experimental bounds on millicharged particles. Therefore
this case does not look very promising.

\item If not all the $U(1)$-charges of the matter fields have the same sign, the $U(1)$ can be non-anomalous
and some matter fields can acquire a non-zero vacuum expectation value partially cancelling the Fayet-Iliopoulos term.
In fact, the scalar potential for the matter fields involves also supersymmetry breaking contributions
to their masses coming from $F$-terms which are generically subleading with respect to the $D$-terms, resulting in a
non exact cancellation of $V_D$. Once the matter fields are integrated out, it turns out that the remaining contribution
to the moduli potential is still dominating the moduli $F$-term potential. Hence we are still facing the dangerous
decompactification problem due to the run-away behaviour of the volume direction.

For compactification manifolds whose volume is controlled by just one 4-cycle like in the isotropic case,
the only way to solve this problem is via fine-tuning by means of warping. Due to the leading order cancellation
in $V_D$, the amount of fine-tuning is less than in the case with no matter fields,
but we would then face the same resulting problems due to the use of warping. On the contrary,
Calabi-Yau three-folds, whose volume is controlled by more than one 4-cycle like in the K3-fibration examples,
look more promising since the dangerous contribution from the scalar potential of the matter fields
could compete against the string loop corrections, resulting in the stabilisation of the K3 divisor and in the
generation of an up-lifting term. In this case we could achieve a Minkowski vacuum without invoking warping
since the fine-tuning could be performed on the coefficients of the $g_s$ corrections.

On top of this, one should check that the Abelian gauge boson which gets a St\"{u}ckelberg mass does not also get a Higgs mass
due to the non-zero VEV of the matter fields, since the contribution from the Higgs mechanism would generically
be the leading effect. We should then envisage a situation similar to the Standard Model where $SU(2)_L\times U(1)\to U(1)_Y$
leaving a massless photon, which in our case would then acquire a mass just via the St\"{u}ckelberg mechanism.
Finally no matter field should violate the experimental bounds on millicharged particles. In order to achieve this,
we should compute the mass of the scalars studying their potential and ensure that our brane set-ups allows the
generation of appropriate Yukawa couplings needed to give a mass to the fermions.
\end{enumerate}
\end{itemize}

\subsubsection{Decompactification problem}

Let us now see why the contribution of $D$-terms from magnetised $D7$-branes
wrapping large 4-cycles, tend to develop a dangerous run-away for the overall volume mode.
We shall first examine the isotropic case following \cite{Burgess:2003ic} and then we shall extend these
results also to the anisotropic one.

\subsubsection*{Isotropic case}

Focusing on the case of a $D7$-brane wrapping $D_b$ with a
non-vanishing gauge flux supported on $t_b$, the general expression (\ref{VD})
for the $D$-term potential takes the form:
\beq
V_D
=\frac{\pi}{\left(\tau_b-f_b q_{bb}/(2g_s)\right)} \left( \sum_j c_{bj} \varphi_j \frac{\partial K}{\partial \varphi_j}
+\frac{q_{bb}}{4\pi}\frac{\partial K}{\partial \tau_b}\right)^2
\simeq \frac{p_1}{\vo^{2/3}} \left( \sum_j c_{bj} \varphi_j \frac{\partial K}{\partial \varphi_j}
-\frac{p_2}{\vo^{2/3}}\right)^2,
\eeq
where $p_1\equiv \pi\left(9\sqrt{2}\right)^{-2/3}$ and $p_2\equiv 9 f_b/\left(6^{1/3}\pi \right)$,
while the volume scaling of the K\"{a}hler potential for the matter fields $\phi_j$
is given by (assuming a diagonal structure) \cite{Conlon:2006tj}:
$K\simeq \sum_j |\phi_j|^2\vo^{-9/4}$. Considering also $F$-term contributions,
the total scalar potential becomes:
\beq
V=V_F+V_D = \frac{p_1}{\vo^{2/3}} \left(
\sum_j c_{bj} \frac{|\phi_j|^2}{\vo^{9/4}}-\frac{p_2}{\vo^{2/3}} \right)^2
+\sum_j k_j \frac{|\phi_j|^2}{\vo^{22/9}}+V_F(T),
\label{Vtotal}
\eeq
where the $k_j$ are $\mc{O}(1)$ numbers and $V_F(T)$ denotes the scalar potential (\ref{Pot}) for the K\"{a}hler moduli.
If the $U(1)$-charges of the matter fields have all an opposite sign with respect to the FI-term,
then clearly the total potential (\ref{Vtotal})
is minimised for $\langle\phi_j\rangle=0$ $\forall j$. Setting each matter field equal to its
vacuum expectation value, we are then left with just the moduli-dependent contribution of the Fayet-Iliopoulos term:
\beq
V = \frac{p}{\vo^2}+V_F(T),\text{ \ \ with \ \ }p=p_1 p_2^2
= \frac{9 f_b^2}{2\pi}.
\label{Vup}
\eeq
Given that (\ref{Pot}) scales as $V_F(T)\simeq \mc{O}(\vo^{-3})$,
we realise that in order to get a Minkowski vacuum
we need to fine-tune $p\simeq \mc{O}(\vo^{-1})$, while from (\ref{Vup})
we notice that $p$ can never be made so small for integer values of the flux coefficient $f_b$.
A possible way-out is to invoke warping but, as we pointed out above, it is hard to
envisage a situation where this can be done without loosing control over the effective field theory.

If some of the matter fields have a $U(1)$-charge with the same sign of the FI-term,
then we can have a leading order cancellation in the $D$-term scalar potential, so
that the $F$-term potential for the matter fields dominates over the contribution from $D$-terms.
Nevertheless we still obtain a run-away for the volume mode since the total potential becomes
(focusing on just one canonically normalised matter field $\varphi_c$):
\beq
V=V_F+V_D = \frac{p_1}{\vo^{2/3}} \left(
c_b |\phi_c|^2-\xi \right)^2+k \frac{|\phi_c|^2}{\vo^2}+V_F(T),\text{ \ \ with \ \ }\xi=\frac{p_2}{\vo^{2/3}}.
\label{NewVtotal}
\eeq
The minimum for the matter field is at:
\beq
\langle|\phi_c|^2\rangle= \frac{\xi}{c_b}-\frac{k}{2c_b^2 p_1 \vo}\simeq \frac{\xi}{c_b},
\eeq
so that, after integrating out $\phi_c$ we are left with:
\beq
V \simeq \frac{k}{c_b}\frac{\xi}{\vo^2}+V_F(T)=\frac{p}{\vo^{8/3}}+V_F(T),
\text{ \ \ with \ \ }p=\frac{9 k f_b}{6^{1/3}\pi c_b}.
\label{NewVup}
\eeq
In this case the fine-tuning needed to obtain a Minkowski vacuum is reduced to
$p\simeq \mc{O}(\vo^{-1/3})$, but (\ref{NewVup}) is showing again that $p$ cannot be
rendered so small since $k$ is an $\mc{O}(1)$ number coming from the computation
of the supersymmetry breaking contribution to the mass of the matter scalars, $f_b$ is
an integer flux-coefficient and $c_b$ is the $U(1)$-charge of the matter fields.

\subsubsection*{Anisotropic case}

This problem subsists also in the anisotropic case.
In fact, focusing on the phenomenologically interesting case of a $D7$-brane wrapping $D_1$ with a
non-vanishing gauge flux supported on $t_2$, the general expression (\ref{VD}) for the $D$-term potential takes the form:
\beq
V_D
=\frac{\pi}{\left(\tau_1-f_2 q_{12}/(2g_s)\right)} \left( \sum_j c_{1j} \,\varphi_j \frac{\partial K}{\partial \varphi_j}
+\frac{q_{12}}{4\pi}\frac{\partial K}{\partial \tau_2}\right)^2
\simeq\frac{\pi}{\tau_1} \left( \sum_j c_{1j}\, \varphi_j \frac{\partial K}{\partial \varphi_j}
-p\frac{\sqrt{\tau_1}}{\vo}\right)^2,
\eeq
where $p\equiv f_2/\left(2\pi\right)$.
The volume scaling of the K\"{a}hler potential for the matter fields $\phi_j$
can be inferred to be (following the same philosophy of \cite{Conlon:2006tj} and assuming a diagonal structure):
$K\simeq \sum_j \tau_1^{1/3}\vo^{-2/3}|\phi_j|^2$. Considering also $F$-term contributions,
the total scalar potential becomes:
\beq
V=V_F+V_D = \frac{\pi}{\tau_1} \left(
\sum_j c_{1j} \tau_1^{1/3}\frac{|\phi_j|^2}{\vo^{2/3}}-p\frac{\sqrt{\tau_1}}{\vo} \right)^2
+\sum_j k_j\tau_1^{1/3}\frac{|\phi_j|^2}{\vo^{8/3}}+V_F(T),
\label{Vtotale}
\eeq
where the $k_j$ are $\mc{O}(1)$ numbers and $V_F(T)$ denotes the scalar potential (\ref{Pot}) for the K\"{a}hler moduli.
If the $U(1)$-charges of the matter fields have all an opposite sign with respect to the FI-term,
then we find $\langle\phi_j\rangle=0$ $\forall j$, leading again to a dangerous run-away behaviour for the volume mode
since the resulting potential is $V = \pi p^2\vo^{-2}+V_F(T)$.

If some of the matter fields have a $U(1)$-charge with the same sign of the FI-term,
then we can have a leading order cancellation in the $D$-term scalar potential, so
that the $F$-term potential for the matter fields dominates over the contribution from $D$-terms.
Nevertheless we still obtain a run-away for the volume mode since the total potential becomes
(focusing on just one canonically normalised matter field $\varphi_c$):
\beq
V=V_F+V_D = \frac{\pi}{\tau_1} \left(
c_1 |\phi_c|^2-\xi_1 \right)^2+k\frac{|\phi_c|^2}{\vo^2}+V_F(T),\text{ \ \ with \ \ }\xi_1=p\frac{\sqrt{\tau_1}}{\vo}.
\label{NewVtotale}
\eeq
The minimum for the matter field is at:
\beq
\langle|\phi_c|^2\rangle= \frac{\xi_1}{c_1}-\frac{k \,\tau_1}{2\pi c_1^2 \vo^2}\simeq \frac{\xi_1}{c_1},
\eeq
so that, after integrating out $\phi_c$ we are left with:
\beq
V \simeq \frac{k}{c_1}\frac{\xi_1}{\vo^2}+V_F(T)=\lambda\frac{\sqrt{\tau_1}}{\vo^3}+V_F(T),
\text{ \ \ with \ \ }\lambda=\frac{k f_2}{2\pi c_1}.
\label{NewVsu}
\eeq
The $\tau_1$-dependent term is still dangerous for the destabilisation of the volume mode since
in order to trust the effective field theory we need to work in the regime $\tau_1\gg 1$.
On top of this, we need now also to fix the K3 divisor $\tau_1$ balancing the term in (\ref{NewVsu})
against other $\tau_1$-dependent contributions to the scalar potential. These can only
arise via string loop corrections to $K$, as we have seen in section \ref{StringLoopCorr},
since the fact that this cycle is non-rigid prevents the presence of non-perturbative effects in $\tau_1$
\footnote{Even taking non-perturbative effects into account, assuming that the deformation moduli
might be fixed by means of non-trivial fluxes, it turns out that no minimum would exist for $\tau_1\gg 1$.}.
Adding to (\ref{NewVsu}) the $g_s$ correction coming from wrapping just a $D7$-brane around
$D_1$, the $\tau_1$-dependent part of the scalar potential looks like:
\beq
V =\left(\lambda\frac{\sqrt{\tau_1}}{\vo}+\frac{A}{\tau_1^2}\right)\frac{W_0^2}{\vo^2},
\text{ \ \ with \ \ }\lambda=\frac{f_2}{6\pi c_1}\text{ \ \ and \ \ }A=\left(g_s C_1^{KK}\right)^2,
\label{Vcombined}
\eeq
where we have substituted in $\lambda$ the correct factor coming from the computation
of the supersymmetry breaking contribution to the mass of the matter scalars: $k= W_0^2/3$.
The potential (\ref{Vcombined}) admits a minimum for $\tau_1$ at:
\beq
\langle\tau_1\rangle =\left(\frac{4 A}{\lambda}\right)^{2/5}\langle\vo\rangle^{2/5}
\,\,\,\,\Leftrightarrow\,\,\,\,\langle\tau_1\rangle =\left(\frac{2 A}{\lambda}\right)^{1/2}\langle\tau_2\rangle^{1/2},
\label{tau1VEV}
\eeq
which implies $\tau_1\ll\tau_2$ in agreement with the anisotropic limit we are interested in.
Substituting (\ref{tau1VEV}) in (\ref{Vcombined}), we end up with a total potential of the form:
\beq
V =\delta\frac{W_0^2}{\vo^{14/5}}+V_F(T),
\text{ \ \ with \ \ }\delta=\frac{5\lambda}{4}\left(\frac{4 A}{\lambda}\right)^{1/5}
\simeq 0.16\cdot \left(\frac{f_2}{c_1}\right)^{4/5}A^{1/5}.
\label{A}
\eeq
If we now integrate out $\tau_3$ from (\ref{Pot}) we are left with a potential for just the volume mode
(setting $\gamma=1$):
\beq
V=\left[-\left(\frac{\ln\vo}{a_3}\right)^{3/2}+\frac{\xi}{g_s^{3/2}}+\frac{4\delta \vo^{1/5}}{3}\right]
\frac{3 W_0^2}{4 \vo^3}.
\eeq
The minimum for $\vo$ is localised at $\langle\vo\rangle\simeq e^{\,a_3 \xi^{2/3}/ g_s}$ and
the requirement of obtaining a vanishing cosmological constant fixes
$\delta=45 \sqrt{\ln\vo}/\left(8 \,a_3^{3/2} \vo^{1/5}\right)\simeq 45 \xi^{1/3}/\left(8\, a_3 g_s^{1/2}\vo^{1/5}\right)$.
Using the expression (\ref{A}) for $\delta$, we see that the level of fine-tuning of the coefficients of
the loop corrections is:
\beq
C_1^{KK}\simeq \frac{7.33\cdot 10^3 \xi^{5/6}}{a_3^{5/2} g_s^{9/4}\vo^{1/2}}\left(\frac{c_1}{f_2}\right)^2
\,\,\,\,\Leftrightarrow\,\,\,\,
\left(\frac{4 A}{\lambda}\right)^{2/5}=\frac{324 \pi^2 c_1^2 \ln\vo}{ f_2^2 a_3^3 \vo^{2/5}}.
\label{Finetune}
\eeq
Substituting this result in (\ref{tau1VEV}) we notice that the VEV of $\tau_1$ is of the order:
\beq
\langle\tau_1\rangle =\frac{324 \pi^2 c_1^2 \ln\langle\vo\rangle}{a_3^3 f_2^2}\simeq
\left(\frac{18 \pi c_1}{a_3 f_2}\right)^2\frac{\xi^{2/3}}{g_s} \gg 1.
\label{VEVtau1}
\eeq
As we shall see in the next section, this case requires always a large amount of fine tuning
to reach the interesting regions of the ($m_{\gamma^\prime}$,$\chi$)-parameter space. Moreover,
it relies on the model-dependent assumption of realising an explicit brane construction
where the hidden photon does not get a large Higgs mass of the order $\langle\varphi_c\rangle\sim \sqrt{\xi_1}\sim M_s$,
and the fermions can evade the stringent bounds on millicharged particles due to the
presence of appropriate Yukawa couplings.

Using similar arguments as above, it can be shown that when we wrap a $D7$-brane also around
$D_2$ with a non-vanishing gauge flux, we face the same destabilisation problem for the volume direction
but without the possibility to fix it by having a leading order cancellation of the $D$-terms as
in the case of a $D7$-brane wrapping $D_1$.

We conclude that $D$-terms associated to $U(1)$ factors living on 4-cycles controlling
the overall volume are the source of a serious run-away problem for the volume mode.
Except for the case of a K3 fibration with a $D7$-brane wrapping $D_1$, where this problem
can be avoided by means of fine-tuning and some model-dependent assumptions, the general
solution seems to rely on the absence of gauge fluxes that prevents the presence of any Fayet-Iliopoulos term,
corresponding to the uninteresting situation of massless hidden photons.

However this is not the end of the story. In fact, in the previous analysis, we
never considered the supersymmetric case of vanishing $D$-terms with zero FI-terms
since, due to the particularly simple form of the overall volume, this situation
would have corresponded to the limit $\vo\to\infty$. In the next section,
we shall therefore solve this problem associated with the $D$-term potential,
focusing on Calabi-Yau three-folds with more complicated topologies where
$\xi=0$ does not correspond to $\vo\to\infty$, but the requirement of having vanishing FI-terms
reduces dynamically the initial compactification manifold to the simple ones we were considering before.

\subsubsection{Vanishing FI-terms with finite volume}

Consider a Calabi-Yau orientifold with $h_{1,1}^+=h_{1,1}$ (i.e. $h_{1,1}^-=0$) even divisors with
$n<h_{1,1}$ of them wrapped by a stack of $D7$-branes with generic gauge fluxes:
\beq
F_2^{D7i}=f^j_{(i)}\hat{D}_j,\,\,\,\,\forall i=1,...,n,
\eeq
that give rise to the following Fayet-Iliopoulos terms:
\beq
\xi_i= \frac{1}{\vo}f_{(i)}^k k_{ijk} t^j =  \frac{1}{\vo}\, q_{ij}\, t^j.
\eeq
It is interesting to notice the nice relation between the $U(1)$-charges $q_{ij}=f_{(i)}^k k_{kij}$
and the chiral intersections:
\begin{eqnarray}
I_{D7i-D7j} &=& \int_{CY} \hat{D}_i \wedge \hat{D}_j \wedge \left(F_2^{D7i}-F_2^{D7j}\right)
=\left(f^k_{(i)}-f^k_{(j)}\right)k_{ijk}= q_{ij} - q_{ji}, \label{ij} \\
I_{D7i-D7i'} &=& 2 \int_{CY} \hat{D}_i \wedge \hat{D}_i \wedge F_2^{D7i}= 2 f^j_{(i)} k_{iij}
=2 q_{ii}, \label{ii'} \\
I_{D7i-D7j'} &=& \int_{CY} \hat{D}_i \wedge \hat{D}_j \wedge \left(F_2^{D7i}+F_2^{D7j}\right)
=\left(f^k_{(i)}+f^k_{(j)}\right)k_{ijk} = q_{ij} + q_{ji}, \label{ij'}
\end{eqnarray}
where we have used the fact that $D_i'=D_i$, while the gauge flux inverts its sign under the orientifold action.
We can then rewrite the Fayet-Iliopoulos terms as a function of the chiral intersections as:
\beq
\xi_i= \frac{1}{\vo}\left[\sum_{j=1}^n\left( \frac{I_{D7i-D7j}+I_{D7i-D7j'}}{2}\right)t_j
+\sum_{j=n+1}^{h_{1,1}} q_{ij} t_j\right],
\label{GenFI}
\eeq
where the first sum is over the 2-cycles dual to the wrapped 4-cycles while the last one is over the unwrapped ones.

If we then impose the absence of chiral intersections between the $D7$-branes and their orientifold images
in order to avoid any problem with millicharged particles, the FI-terms (\ref{GenFI}) simplify to:
\beq
\xi_i= \frac{1}{\vo}\sum_{j=n+1}^{h_{1,1}} q_{ij} t_j,
\eeq
with an explicit dependence only on the 2-cycles dual to the 4-cycles not wrapped by any brane.

Focusing on the supersymmetric locus corresponding to vanishing FI-terms, we then obtain
a homogeneous system of $n$ linear equations in $m\leq (h_{1,1}-n)$ unknowns
(we have $m< (h_{1,1}-n)$ if some unwrapped 4-cycles do not intersect any of the wrapped ones) which, in the
case of linearly independent equations, admits only the trivial solution if $m\leq n$.

We shall therefore focus on the case $m=(h_{1,1}-n)$ $\Leftrightarrow$ $n<h_{1,1}\leq 2 n$,
and look for Calabi-Yau geometries that
admit a singular limit to our previous examples, dynamically driven by the supersymmetric requirement
of having vanishing $D$-terms. Moreover the absence of chiral matter renders our results truly
model independent and the resulting improved control over moduli stabilisation strengthens the
robustness of our predictions.

\subsubsection*{Isotropic case}

In this section we shall show how the supersymmetric requirement of having vanishing
$D$-terms allows us to obtain isotropic compactifications
with just one large 4-cycle controlling the overall volume,
as a singular limit of different Calabi-Yau three-folds
with more complicated topologies.

We shall start from the same manifold discussed in section \ref{AnisComp},
that is an orientifold of the Calabi-Yau given by the degree 12
hyper-surface embedded in $\mathbb{C}P^4_{[1,1,2,2,6]}$. The volume can be expressed in
terms of the 2-cycle moduli as in (\ref{Vol11226}) or as a function of the 4-cycle moduli
as in (\ref{vol11226}) with the only two non-zero intersection numbers given by $k_{122}=2$ and $k_{222}=4$.

We now wrap a single $D7$-brane around the divisor $D_2$ whose volume
is $\tau_2=2 t_2\left(t_1+t_2\right)$ turning on also a generic gauge flux on this brane:
$F_2= f_1 \hat{D}_1+ f_2 \hat{D}_2$. On the other hand, we do not wrap any $D7$-brane
around the other divisor $D_1$ with volume given by $\tau_1=t_2^2$.

The requirement of avoiding the generation of chiral matter at the intersection between
the $D7$-brane and its orientifold image, constraints the form of the integer flux coefficients.
In fact, the number of chiral bi-fundamental states at this intersection reads:
\beq
I_{D7-D7'} = 2 \int_{CY} \hat{D}_2 \wedge \hat{D}_2 \wedge F_2= 4\left( f_1 +2 f_2\right)=0
\,\,\,\,\Leftrightarrow\,\,\,\,  f_1= -2 f_2,
\label{chiralInt}
\eeq
where we have used the fact that we are dealing with even 4-cycles under the orientifold, i.e. $D_1'=D_1$,
while the gauge flux flips sign: $F_2 \to - F_2$. Recalling the general relation between the $U(1)$-charges
and the Fayet-Iliopoulos terms,
we find that the only non-vanishing charge is $q_{21}=2 f_2$ leading to:
\beq
\xi_2 = \frac{q_{2j}t_j}{\vo}=\frac{ 2 f_2 t_1}{\vo}.
\eeq
The supersymmetry requirement of having vanishing $D$-terms then forces the 2-cycle
$t_1$ to shrink to zero size: $t_1\to 0$. In this singular limit, the initial
Calabi-Yau takes exactly the same form of the isotropic case studied in section \ref{IsComp}
while all the divisors stay finite:
\beq
\vo = \left(t_1 +\frac{2}{3}t_2\right)t_2^2 \to \frac{2}{3}t_2^3 =
\frac{2}{3}\,\tau_1^{3/2} \,\,\,\,
\Leftrightarrow \,\,\,\,\tau_1 = t_2^2, \,\,\tau_2 = 2 t_2\left(t_1+t_2\right) \to 2 t_2^2 = 2 \tau_1.
\eeq
Therefore in this case the $D$-term potential is under control and the $U(1)$ gauge boson living on $D_2$
can become massive via the Green-Schwarz mechanism due to its coupling to a particular combination of the canonically
normalised two-forms $\mc{D}_2^1$ and $\mc{D}_2^2$. In order to compute the mass of the hidden
photon we need the diagonalising matrix evaluated at $\tau_2 = 2 \tau_1$ which looks like:
\beq
 C^i_j=\frac{1}{\tau_1}\left(
 \begin{array}{cccc}
 c_1 && c_3 \\
 c_2   &&  c_4
 \end{array}
 \right),\text{ \ with \ }c_{1,2}=(-1\mp\sqrt{17}) \sqrt{\frac{3}{2 (51\mp5 \sqrt{17})}},
 \text{ \ and \ }c_{3,4}=2 \sqrt{\frac{6}{51\mp 5 \sqrt{17}}}. \notag
\eeq
Therefore the mass of the hidden photon living on $D_2$ turns out to be:
\beq
m_{\gamma^\prime}=\sqrt{M_{21}^2+M_{22}^2}
\sim \, f_2 \frac{M_P}{\tau_1^{3/2}}
\sim f_2 \frac{M_P}{\tau_1^{3/2}}, \eeq
which is exactly of the same form as (\ref{Mbb}) once we identify $\tau_1$ with the
big 4-cycle $\tau_b$.

\subsubsection*{Anisotropic case}

Start from a Calabi-Yau with three K\"{a}hler moduli and volume of the form:
\beq
\vo=t_1 t_2 \left(t_2 + t_3\right),
\eeq
so that the only non-zero intersection numbers are $k_{122}=2$ and $k_{123}=1$.
It can be checked that the Hessian $\partial^2 \vo/\left(\partial t_i \partial t_j\right)$
admits one positive and two negative eigenvalues in accord with the generic property of
Calabi-Yau manifolds that requires the signature of the Hessian to be $(1,h_{1,1}-1)$.

The 4-cycle moduli are given by:
\beq
\tau_1=t_2\left(t_2+t_3\right),\text{ \ \ \ \ }\tau_2=t_1\left(2 t_2+t_3\right),
\text{ \ \ \ \ }\tau_3=t_1 t_2,
\eeq
and the volume can be reexpressed in terms of them as:
\beq
\vo=\sqrt{\tau_1\tau_3\left(\tau_2-\tau_3\right)}.
\eeq

We now wrap a single $D7$-brane both around the divisor $D_1$ and $D_2$. In addition
we turn on generic gauge fluxes on these branes:
\beq
F_2^{D71}= f_2 \hat{D}_2+ f_3 \hat{D}_3, \text{ \ \ and \ \ }
F_2^{D72}= g_1 \hat{D}_1 + g_2 \hat{D}_2+ g_3 \hat{D}_3,
\eeq
with the additional constraint that no chiral matter is generated at each of the
possible intersections between branes and their orientifold images:
\begin{eqnarray}
I_{D71-D72} &=& \int_{CY} \hat{D}_1 \wedge \hat{D}_2 \wedge \left(F_2^{D71}-F_2^{D72}\right)
=2\left(f_2-g_2\right) + f_3-g_3=0 \label{12} \\
I_{D71-D71'} &=& 2 \int_{CY} \hat{D}_1 \wedge \hat{D}_1 \wedge F_2^{D71}=0 \label{11'} \\
I_{D71-D72'} &=& \int_{CY} \hat{D}_1 \wedge \hat{D}_2 \wedge \left(F_2^{D71}+F_2^{D72}\right)
=2\left(f_2+g_2\right) + f_3+g_3=0 \label{12'} \\
I_{D72-D72'} &=& 2 \int_{CY} \hat{D}_2 \wedge \hat{D}_2 \wedge F_2^{D72}=4 g_1=0 \label{22'} \\
I_{D72-D71'} &=& I_{D71-D72'}, \text{ \ \ \ \ }I_{D72'-D71'} = - I_{D71-D72}.
\end{eqnarray}
We remind the reader that in the previous expressions we have used the fact that we are dealing
with even 4-cycles under the orientifold, i.e. $D_i'=D_i$, while the gauge flux flips sign:
$F_2^i \to - F_2^i$. The combined constraints (\ref{12}) and (\ref{12'}) imply $f_3 = -2 f_2$
and $g_3 = -2 g_2$, while (\ref{11'}) is satisfied by construction since the divisor $D_1$ has
no self-intersection, i.e. $k_{1ij}=0$ $\forall i,j$, and (\ref{22'}) forces $g_1=0$.

Recalling the general relation between the $U(1)$-charges and the Fayet-Iliopoulos terms,
we find that the only non-vanishing charge is $q_{13}=f_2$ leading to:
\beq
\xi_1 = \frac{q_{1j}t_j}{\vo}=\frac{f_2 t_3}{\vo},\text{ \ \ \ and \ \ \ }
\xi_2 = \frac{q_{2j}t_j}{\vo}=0.
\eeq
The supersymmetry requirement of having vanishing $D$-terms then forces the 2-cycle
$t_3$ to shrink to zero size: $t_3\to 0$. In this singular limit, the initial
Calabi-Yau takes exactly the same form of the K3 fibration studied in the previous section
while all the divisors stay finite:
\beq
\vo \to  t_1 t_2^2 = \frac{1}{2}\sqrt{\tau_1}\tau_2 \,\,\,\,
\Leftrightarrow \,\,\,\,\tau_1 \to t_2^2, \,\,\tau_2 \to 2 t_1 t_2,
\,\,\tau_3 \to t_1 t_2 = 2 \tau_2.
\eeq
Therefore in this case the $D$-term potential is under control and the brane set-up
is the right one to generate the string loop correction to the K\"{a}hler potential
needed to fix the K3 divisor.

However in the previous sections, we have seen that the $U(1)$ gauge boson living on $D_1$
becomes massive due to its coupling with the two-form $\mc{D}_2^2$ while in our case
$q_{12}=0$ and $F_2^{D71}$ couples to a particular combination of the canonically
normalised two-forms $\mc{D}_2^2$ and $\mc{D}_2^3$. However the final mass of the hidden
photon takes the same form since the diagonalising matrix evaluated at $\tau_3 = 2 \tau_2$
looks like:
\beq
 C^i_j=\frac{1}{\tau_2}\left(
 \begin{array}{ccccc}
 \tau_2/\tau_1 && 0 && 0 \\
 0 && \lambda_1 && \lambda_3 \\
 0 && \lambda_2   &&  \lambda_4
 \end{array}
 \right),\text{ \ with \ }\lambda_{1,2}=\frac{1\mp \sqrt{5}}{5^{1/4}\sqrt{\sqrt{5}\mp 2}},
 \text{ \ and \ }\lambda_{3,4}=\frac{2}{5^{1/4}\sqrt{\sqrt{5}\mp 2}}. \notag
\eeq
Therefore the mass of the hidden photon living on $D_1$ turns out to be:
\beq
m_{\gamma^\prime}^{D71}=\sqrt{M_{12}^2+M_{13}^2}=
\sqrt{\frac{\left(\lambda_2^2 + \lambda_4^2\right)}{4\pi^2}} \, f_2 \frac{M_P}{\sqrt{\tau_1}\tau_2}
\sim f_2  \frac{M_P}{\sqrt{\tau_1}\tau_2}, \eeq
which is exactly of the same form as (\ref{M12}).
We finally point out that the hidden photon living on $D_2$ is exactly massless since
the corresponding FI-term is vanishing, $\xi_2=0$, due to the requirement of avoiding
chiral intersections and the particularly simple structure of the intersection numbers.

\section{Phenomenological implications}
\label{SECTION:IMPLICATIONS}

Let us focus on the phenomenologically interesting case of K3 fibrations
with a $D7$-brane wrapped around $D_1$ with $f_2$ units of gauge flux on $t_2$.
We have seen that there are two ways to stabilise the K3 divisor either via $g_s$
corrections to $K$ for vanishing FI-terms or by making these corrections compete
with $D$-terms for non-zero FI-terms. Let now examine the predictions for the features
of the hidden photon in these two separate scenarios.

\subsection{Vanishing FI-term}

\begin{figure}
\begin{center}
\includegraphics[width=0.95\textwidth]{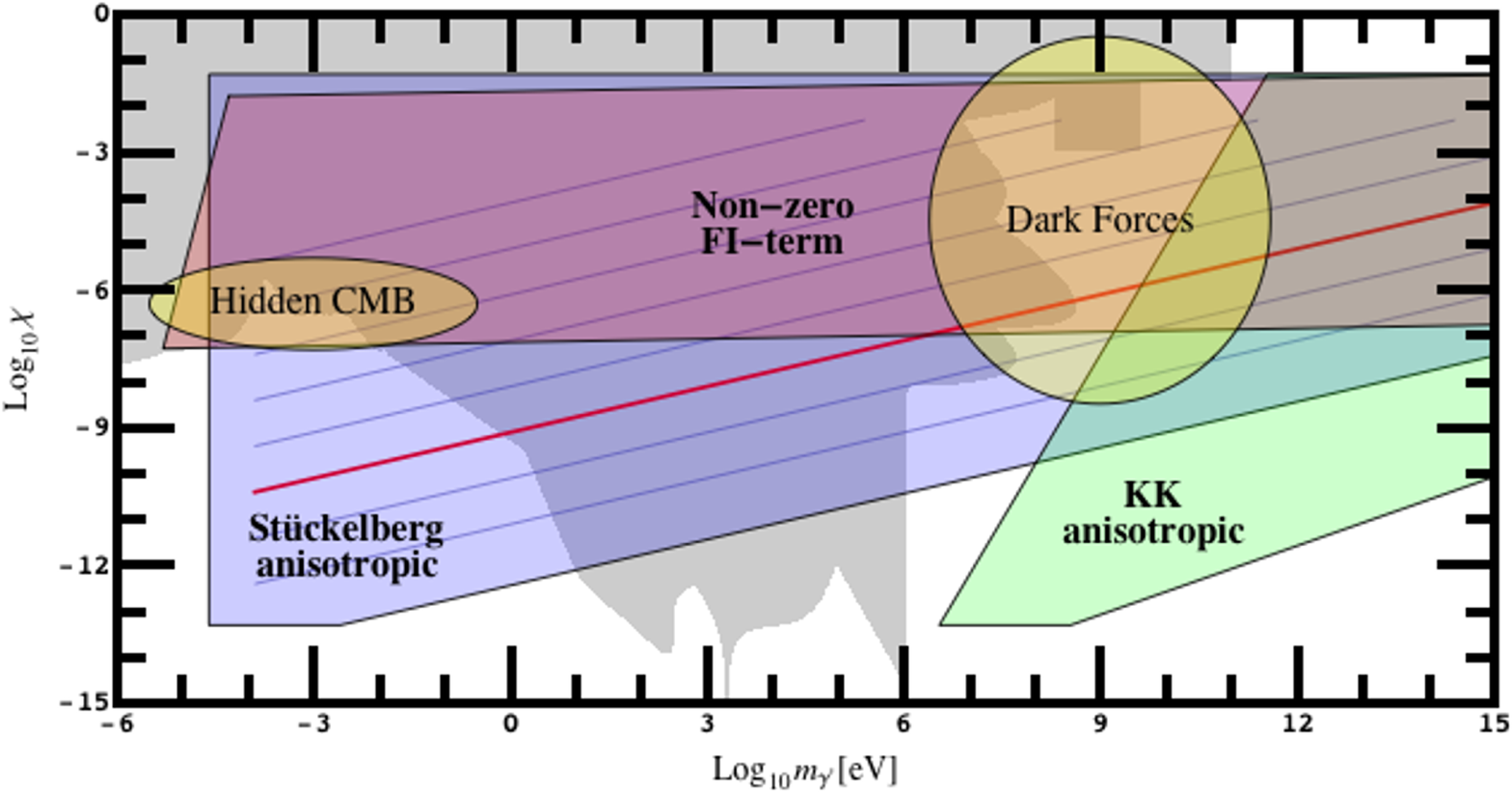}
\end{center}
\vspace{-2.ex}
\caption{
Predictions from anisotropic compactifications. The area predicted by models with vanishing FI-terms is shown
in light blue and marked as ``St\"uckelberg anisotropic''. The lines denote different values of from bottom to top,
$\kappa=1,10^{-3},10^{-6}\ldots10^{-21}$.
The red line denotes a natural $\kappa=10^{-6}$. The green area marked ``KK anisotropic'' denotes the region where we expect the corresponding Kaluza-Klein modes.
Finally the light red area ``Non-zero FI-terms'' corresponds to parameter values expected in models with non-vanishing Fayet-Iliopoulos terms.
The existing experimental and observational constraints are marked in grey.
As in Fig.~1
we have marked phenomenologically interesting areas in yellow.
}
\label{predictions}
\end{figure}

Fixing $\tau_1$ via string loop corrections to the K\"{a}hler potential we can rewrite
the relation (\ref{MchiRel}) between $m_{\gamma^\prime}$ and $\chi$ as:
\beq
\frac{\chi}{10^{-8}} \sim \kappa^{-1/3}  \left(\frac{m_{\gamma^\prime}}{\text{\ GeV}} \right)^{1/3} \qquad \Leftrightarrow \qquad \kappa \sim \left(\frac{10^{-8}}{\chi}\right)^3 \left(\frac{m_{\gamma^\prime}}{\text{\ GeV}} \right),
\eeq
where $\kappa$ is a free parameter that is naturally small since it is proportional to $g_s^2\ll 1$:
\beq
\langle \tau_1 \rangle =  \kappa \,\langle\tau_2\rangle,
\text{ \ \ \ \ with \ \ \ \ }\kappa\equiv \frac{\left(g_s C_1^{KK}\right)^2}{C_{12}^W}.
\eeq
The masses and mixings reachable in this type of setup are shown in Fig.~\ref{predictions} as the light blue area. For large volumes of the cycle supporting the hidden photon we typically also obtain fairly light KK modes of the hidden photon. Assuming that their mixing with the zero mode of the electromagnetic field is of similar size we expect values in the light green area of Fig.~\ref{predictions}.

Let us show some interesting values:
\begin{itemize}
\item $\chi \sim 10^{-6}$ gives $m_{\gamma^\prime}\sim \kappa\, 10^6$ GeV and we obtain:
\begin{enumerate}
\item Dark forces: $m_{\gamma^\prime}\sim 1$ GeV for $\kappa \sim 10^{-6}$,

\item Hidden CMB: $m_{\gamma^\prime}\sim 1$ meV for $\kappa \sim 10^{-18}$,
\end{enumerate}

\item $\chi \sim 10^{-7}$ gives $m_{\gamma^\prime}\sim \kappa\, 10^3$ GeV and we obtain:
\begin{enumerate}
\item Dark forces: $m_{\gamma^\prime}\sim 1$ GeV for $\kappa \sim 10^{-3}$,

\item Hidden CMB: $m_{\gamma^\prime}\sim 1$ meV for $\kappa \sim 10^{-15}$,
\end{enumerate}
\end{itemize}
Therefore we realise that we can reach the dark force regime naturally while we
need some amount of fine-tuning to allow for the presence of a hidden CMB.

Let us now check the actual amount of fine tuning and the corresponding value of
the overall volume which sets all the fundamental scales in our theory. To do that let us recall
that naturally we have $g_s\sim 0.1$. Moreover,  $C_1^{KK}$ and $C_{12}^W$ are unknown functions 
of the complex structure moduli that can, in principle, be tuned by an appropriate choice of fluxes, however the natural expectation is $C_1^{KK}\sim C_{12}^W \sim \mc{O}(1)$. Deviations from these natural values require a certain amount of fine-tuning.

\subsubsection*{Natural Dark Forces for intermediate scale strings}

\begin{itemize}
\item $\kappa =2.5\cdot 10^{-6}\sim 10^{-6}$ can be obtained choosing $g_s=0.1$, $C_1^{KK}=0.05$ and
$C_{12}^W=0.1$ corresponding to a kinetic mixing parameter of the order $\chi \sim 10^{-6}$. The
VEVs of the two moduli become $\tau_1\sim 10^{-4} \chi^{-2}\sim 10^8$ and
$\tau_2=\tau_1/\kappa\sim 10^{14}\gg \tau_1$ leading to a volume of the order
$\vo \simeq \alpha\sqrt{\tau_1}\tau_2 \sim 10^{17}$ for $\alpha=0.1$. Thus the
string scale turns out to be intermediate $M_s\sim 10^{11}$ GeV.

\item $\kappa = 10^{-3}$ can be obtained choosing $g_s=C_1^{KK}=C_{12}^W=0.1$
corresponding to a kinetic mixing parameter of the order $\chi \sim 10^{-7}$. The
VEVs of the two moduli become $\tau_1\sim 10^{-4} \chi^{-2}\sim 10^{10}$ and
$\tau_2=\tau_1/\kappa\sim 10^{13}\gg \tau_1$ leading again to a volume of the order
$\vo \simeq \alpha\sqrt{\tau_1}\tau_2 \sim 10^{17}$ for $\alpha=0.1$ together with an intermediate string scale.
\end{itemize}
We therefore conclude that we can obtain dark forces for natural values of the underlying
parameters in scenarios where the string scale is intermediate. These scenarios
are favoured also by the fact that TeV-scale supersymmetry can be achieved since the soft masses
scale as $M_{soft}\sim W_0 M_P /\vo \sim 1$ TeV for $W_0\sim 40$. Moreover $M_s\sim 10^{11}$ GeV
yields also a decay constant for the QCD axion in the allowed region
and the right Majorana mass scale for right handed neutrinos.

We finally notice that the dark force case corresponds to Calabi-Yau geometries with a slight anisotropy
since there is only a mild hierarchy between the characteristic size of the base $L\sim t_1^{1/2}= \sqrt{\vo/\tau_1}\sim 10^4$
and that of the K3 fibre $l\sim\tau_1^{1/4}\sim 10^2$.

\subsubsection*{Hidden CMB with KK Dark Forces and strings at the LHC}

\begin{itemize}
\item $\kappa = 10^{-15}$ can be obtained choosing $g_s=0.01$, $C_1^{KK}\sim 10^{-4}$ and
$C_{12}^W\sim 10^3$ corresponding to a kinetic mixing parameter of the order $\chi \sim 10^{-7}$.
The VEVs of the two moduli become $\tau_1\sim 10^{-4} \chi^{-2}\sim 10^{10}$ and
$\tau_2=\tau_1/\kappa\sim 10^{25}\gg \tau_1$ leading to a volume of the order
$\vo \simeq \alpha\sqrt{\tau_1}\tau_2 \sim 10^{30}$ for $\alpha=1$.
Thus we are in the extreme case of TeV-scale strings: $M_s\sim 1$ TeV.

\item $\kappa = 10^{-18}$ can be obtained choosing $g_s=10^{-3}$, $C_1^{KK}=10^{-4}$ and $C_{12}^W=10^4$
corresponding to a kinetic mixing parameter of the order $\chi \sim 10^{-6}$. The
VEVs of the two moduli become $\tau_1\sim 10^{-4} \chi^{-2}\sim 10^8$ and
$\tau_2=\tau_1/\kappa\sim 10^{26}\gg \tau_1$ leading again to a volume of the order
$\vo \simeq \alpha\sqrt{\tau_1}\tau_2 \sim 10^{30}$ for $\alpha=1$ together with TeV-scale strings.
\end{itemize}
We therefore conclude that we can obtain a hidden CMB candidate 
by fine-tuning the values of the underlying
parameters in
scenarios with $M_s\sim 1$ TeV. These scenarios
are very promising from several other points of view: they provide a solution to
the hierarchy problem that does not rely on supersymmetry, and they might shed new light on the solution of the cosmological constant problem \cite{ADDfromStrings}.

Furthermore they can be detected in the lab by four different means: via string resonances and deviations from Standard Model quark scattering at the LHC; at light shining through a wall experiments such as ALPS; they lead to
large extra dimensions and light moduli mediating long range fifth forces
that would give rise to modifications of Newton's law at the edge of
detectability; and the Kaluza-Klein excitations of the hidden gauge bosons are in the Dark Forces regime and could thus be produced in the next generation of experiments searching for these.

We finally notice that the hidden CMB case corresponds to Calabi-Yau three-fold with a very anisotropic shape
since there is a large hierarchy between the characteristic size of the base $L\sim t_1^{1/2}= \sqrt{\vo/\tau_1}\sim 10^{11}$
and that of the K3 fibre $l\sim\tau_1^{1/4}\sim 10^2$.

\subsection{Non-zero FI-term}

Fixing $\tau_1$ via the interplay of $D$-terms and
string loop corrections to the K\"{a}hler potential we can rewrite
the relation (\ref{MchiRel}) between $m_{\gamma^\prime}$ and $\chi$ as:
\beq
m_{\gamma^\prime}\simeq  10^{20} \alpha f_2 \frac{\sqrt{\tau_1}}{\vo}\chi \text{\ GeV}
\simeq 5\cdot 10^{17} \alpha \frac{f_2}{\vo} \text{\ GeV}.
\label{Rel}
\eeq
Substituting in (\ref{Rel}) the expression (\ref{VEVtau1}) for the VEV of $\tau_1$ in terms of $\chi$, we
end up with:
\beq
m_{\gamma^\prime}\simeq 5\cdot 10^{17}\mu_1\, e^{-\mu_2 \chi^{-2}} \text{\ GeV},
\eeq
where $\mu_1=\alpha f_2$ and $\mu_2=25\cdot 10^{-6}\sigma^{-1}$, with $\sigma$ that is a free parameter
whose value is fixed by the requirement of getting a Minkowski vacuum:
\beq
\langle\tau_1\rangle =\sigma \ln\langle\vo\rangle,\text{ \ \ \ \ with \ \ \ \ }
\sigma\equiv\frac{324 \pi^2 c_1^2 }{a_3^3 f_2^2}.
\eeq
Varying the integers in this equation within reasonable limits ($1-10$) we can reach values within the red area of Fig.~\ref{predictions}.

Let us now illustrate the phenomenological implications of these results with the help of
some parameter fits that lead to $\chi\simeq 10^{-6}$.
The dark force case with $m_{\gamma^\prime}\sim 1$ GeV can be achieved for $\mu_1 =0.1$ and $\mu_2=4.1565\cdot 10^{-11}$.
These two values can be obtained choosing $\alpha=0.1$, $f_2=1$, $c_1=8$ and
$a_3=2\pi/N_3$ with $N_3=9$. The VEV of the K3 divisor becomes
$\tau_1\simeq 25\cdot 10^{-6} \chi^{-2}=2.5\cdot 10^7$ while the volume is of the order
$\vo \simeq e^{\tau_1/\sigma} \sim 10^{18}$ corresponding to an intermediate string scale.
Given that the volume can be also expressed as $a_3\ln\vo\simeq \xi^{2/3}/g_s$, we can choose
$\xi=1.5$ and $g_s=0.045$. Using (\ref{Finetune}), we can finally check that
no fine-tuning of the string loop corrections is needed in order to get a vanishing cosmological constant
since the above choice of parameters sets $C_1^{KK}\simeq 1.63$.

We therefore conclude that, even in this case where the K3 divisor is fixed
by the interplay of $D$-terms and $g_s$ corrections, we can obtain dark forces
for natural values of the underlying parameters in scenarios where the string scale is intermediate.
However this case looks less promising than the one with vanishing FI-terms since
the additional constaint coming from the requirement of a viable up-lifting reduces
the reliability of our predictions. In fact, due to the exponential dependence of $m_{\gamma^\prime}$ on $\chi$,
a small change in our choice of parameters gives drastic changes for the mass of the hidden photon
at fixed kinetic mixing. For example, if we just change $c_1=8$ to $c_1=7$ in the above fit,
the prediction for the $U(1)$ mass gets modified from $m_{\gamma^\prime}\simeq 1$ GeV to $m_{\gamma^\prime}\simeq 10^{-6}$ GeV.

With other small changes, like $\alpha=0.1 \to 10$ and $c_1=8 \to 6$,
we can easily reach the interesting hidden CMB regime for $m_{\gamma^\prime}\sim 1$ meV.
The VEV of the K3 divisor is still of the order $\tau_1\simeq 10^7$ while the volume
now becomes $\vo \simeq e^{\tau_1/\sigma} \sim 10^{32}$ corresponding to the extreme case of TeV-scale strings.
Such a large value of $\vo$ can be obtained for $\xi=1.5$ and $g_s=0.025$. Using (\ref{Finetune}),
we can finally check that, contrary to the dark force case, the upliting now requires a large fine-tuning
of the coefficient of the loop corrections of the order $C_1^{KK}\simeq 3.3\cdot 10^{-7}$.

\section{Conclusions}
\label{SECTION:CONCLUSIONS}

We have shown that allowing for anisotropy in LARGE volume compactifications greatly enhances the phenomenological possibilities for hidden D-brane $U(1)$s. In this case, in addition to  collapsed, small, or hyperweak cycles, it is possible to wrap (hidden) branes on  ``milliweak'' cycles. Each of these will give different ranges of gauge couplings and thus kinetic mixing with the hypercharge, but since in this case the St\"uckelberg masses of the $U(1)$s become more weakly correlated with the volume of the cycle, a milliweak cycle allows for the attractive possibility of a very small mass but moderate (and thus observable) mixing.

In fact, naively there is an embarrassment of riches; the possible masses and mixings become so diverse as to render the scenario almost unpredictive, with the exception of lower bounds due to there being a maximum volume of the compactification (of $\V \sim 10^{30}$ since the string scale cannot be below $\mathcal{O}(\mathrm{TeV})$). We therefore considered the constraints imposed by insisting on moduli stabilisation, taking careful account of the relationship between fluxes required to give the $U(1)$s masses and the presence of D-terms and hidden chiral matter. These constraints then translate into requirements on uncalculable parameters in the model such as the coefficients of the loop corrections to the K\"ahler potential. We found that without any fine-tuning it is possible to have hidden $U(1)$s in the Dark Forces regime even for intermediate scale strings, and so we could soon be probing intermediate scale string effects in the lab!

We also found that we can realise the ``hidden CMB'' scenario of a hidden $U(1)$ of mass $\sim$ meV and mixing $\sim 10^{-6}$, with the price being some fine tuning. In compensation we surprisingly find multiple ways to test it: other than cosmological observations, it can be directly tested in lab experiments at very low energies, in Dark Forces experiments due to the hidden Kaluza-Klein modes, and at the LHC since the string scale must be low.

We hope that we have provided ample motivation and tools to search for these setups in more complete models, including explicit brane constructions with tadpole cancellation. The reward for this endeavour would be a way to probe in the lab the hidden sectors that generically arise, and interesting hidden sector model building.

\section{Acknowledgements}

We would like to thank Fernando Quevedo for useful discussions, and the organisers of the XXIII Bad Honnef ``Beyond the Standard Model'' workshop for hospitality while this paper was in the final stages of preparation. 
MDG is supported by the German Science Foundation (DFG) under SFB 676.

\appendix

\section{$U(1)$ masses from dimensional reduction}
\label{APPENDIX:Background}

\subsection{U(1) factors from D-branes}

A very important ingredient of Calabi-Yau flux compactifications
is the presence of $Dp$-branes which wrap internal $(p-3)$-cycles
and have to fill the four-dimensional space-time in order not to break Poincar\'{e}
invariance. Each space-time filling $Dp$-brane comes along with a
$U(1)$ gauge theory that lives on its world volume.
Thus string compactifications naturally come along with many $U(1)$ gauge bosons.

The dynamics of a $Dp$-brane is governed by the Dirac-Born-Infeld
action $S_{DBI}$ together with a Chern-Simons action $S_{CS}$:
\begin{eqnarray}
S_{DBI}&=&-\mu_{p} e^{-\phi}\int_{\mc{W}}d^{p+1}\xi
\sqrt{-\textrm{det}\left[\imath^*(G+B_2)+l_s^2 F_2/(2\pi)\right]}, \label{DBI} \\
S_{CS}&=&\mu_{p}e^{-\phi}\int_{\mc{W}}\sum_p \imath^* \left(C_p \right)\wedge e^{\imath^* (B_2)+l_s^2 F_2/(2\pi)}, \label{CS}
\end{eqnarray}
where $\phi$ is the dilaton, $\mu_{p}$ is the tension of the $Dp$-brane which is equal to
its RR-charge since the $Dp$-brane has to satisfy a BPS condition,
$G$ denotes the 10D metric, $B_2$ is the NS-NS 2-form, $F_2$ is the gauge field strength,
and $C_p$ is a R-R $p$-form. The integrals in (\ref{DBI}) and (\ref{CS}) are taken over the
$(p+1)$-dimensional world-volume $\mc{W}$ of the $Dp$-brane, which
is embedded in the ten dimensional space-time manifold
$\mathcal{X}_{10}=\mathbb{R}^{3,1}\times \mathcal{M}_6$,
where $\mc{M}_6$ is a 6D Calabi-Yau manifold, via the map $\imath:\mc{W}\hookrightarrow
\mathcal{X}_{10}$. $\imath^*$ denotes the pullback operation.

From now on, we shall focus on type IIB flux compactifications since
this is the context where moduli stabilisation is best understood.
We shall also be interested in the case of a $D7$-brane wrapping an internal
4-cycle $D$ which is a smooth divisor of the Calabi-Yau three-fold.
The volume of a generic 4-cycle $D_i$ is given by the real part of the K\"{a}hler modulus $T_i$
which in 4D Einstein-frame is defined as:
\begin{equation}
T_i \equiv \left( \int_{D_i}
\sqrt{g}\, d^4 y + i \int_{D_i} C_4 \right) \frac{e^{-\phi}}{l_s^4} \equiv \tau_i +i b_i,\,\,\,\,\,\,\,\,\,i=1,...,h_{1,1},
\label{defT}
\end{equation}
where $h_{1,1}$ is one of the Calabi-Yau Hodge numbers and $C_4$ is the 10D R-R 4-form.

The standard Maxwell action can be obtained from expanding the DBI action (\ref{DBI})
in powers of the field strength\footnote{We are neglecting the background NSNS two-form $B_2$ since we shall look at orientifold
projections such that $h_{1,1}^-=0$.}:
\begin{eqnarray}
S_{DBI}&=&-\mu_7 e^{-\phi}\int_{\mathbb{R}^{3,1}\times D_i}d^8\xi
\sqrt{-\textrm{det}\left[\varphi^*(G)+l_s^2 F_2/(2\pi)\right]} \label{DBIS} \\
&=&-\mu_7 e^{-\phi}\int_{\mathbb{R}^{3,1}\times D_i}d^8\xi
\sqrt{-\textrm{det}\left[\varphi^*(G)\right]}\left(1+ \frac{l_s^4}{16 \pi^2}F_{MN}F^{MN}
-\frac{l_s^8}{128\pi^4}\left(F_{MN}F^{MN}\right)^2+...\right), \notag
\end{eqnarray}
and then performing the dimensional reduction from 8D to 4D:
\beq
-\frac{\mu_7 e^{-\phi} l_s^4}{16 \pi^2}\int_{\mathbb{R}^{3,1}\times D_i}d^8\xi
\sqrt{-\textrm{det}\left[\varphi^*(G)\right]}F_{MN}F^{MN} \,\,\,\,\rightarrow \,\,\,\,
-\left(\frac{\mu_7 l_s^8}{16 \pi^2}\right) \tau_i
\int_{\mathbb{R}^{3,1}} F_{\mu\nu}F^{\mu\nu} d^4 x. \notag
\eeq
The $D7$-brane tension is given by $\mu_7= 2\pi/(g_s l_s^8)$, and so we obtain the final result:
\beq
\mathcal{L}_{kin}=-\frac{1}{4 g_i^2}
\int_{\mathbb{R}^{3,1}} F_{\mu\nu}F^{\mu\nu} d^4 x \,\,\,\,\,\,\text{with}\,\,\,\,\,\,g_i^2=\frac{2\pi}{\tau_i}.
\label{FF}
\eeq

\subsection{Massive U(1)s from internal fluxes}\label{stueckel}

A $U(1)$ gauge boson living on a $D7$-brane can acquire a mass by turning on
an internal magnetic flux in the world-volume of the $D7$-brane.
In fact, turning on a 2-form gauge flux on a 2-cycle internal to the 4-cycle
wrapped by the $D7$ generates a coupling between the $U(1)$ gauge boson and the K\"{a}hler
modulus (\ref{defT}) corresponding to the 4-cycle Poincar\'e dual to the 2-cycle supporting
the magnetic flux. Then the axion, that is the imaginary part of the charged K\"{a}hler modulus,
gets eaten up by the $U(1)$ gauge boson which
becomes massive via the St\"{u}ckelberg mechanism. This is the way in which
string theory cures any problem coming from dangerous anomalous $U(1)$s which
acquire $\mc{O}(M_s)$-masses through the Green-Schwarz mechanism,
and so they disappear from the 4D effective field theory.
However we shall be interested in non-anomalous $U(1)$s which can still
become massive via the same mechanism, but their mass can be much lighter
than $M_s$. Therefore these hidden photons have to be included in
the description of the 4D effective field theory coming from string compactifications.

Let us see more in detail how this happens. The expansion of the Chern-Simons action (\ref{CS})
contains a coupling with the 10D R-R 4-form $C_4$ which looks like:
\beq
\C{L} \supset - 2\pi \frac{e^{-\phi}}{l_s^2} \int_{\mathbb{R}^{3,1} \times D_i}
\frac{F_2}{2\pi} \wedge C_4 \wedge \frac{F_2}{2\pi}.
\eeq
The R-R form $C_4$ can be decomposed as:
\beq
C_4 = Q_2^i(x) \wedge \hat{D}_i(y) + b_i(x) \tilde{D}^i(y),\text{ \ \ }i=1,...,h_{1,1},
\label{defC4}
\eeq
where the $\tilde{D}_i$ are a basis of harmonic (2,2)-forms of $H_{2,2}(\mc{M}_6)$,
dual to the (1,1)-forms $\hat{D}_i$, while the 4D fields $b_i(x)$ are the axions defined in (\ref{defT})
and $Q_2^i(x)$ are 2-forms dual to the $b_i (x)$ (due to the self-duality of $F_5 = dC_4 = \star_{10D} F_5$).
Taking both of the $F$'s to be with non-compact indices, and reducing $C_4$ along the divisor $D_i$,
gives rise to the axion-dependent CP-odd coupling:
\beq
\C{L} \supset -\frac{e^{-\phi}}{2 \pi l_s^4} \left(\int_{D_i} C_4 \right) \int_{\mathbb{R}^{3,1}} F_2 \wedge F_2
= \frac{b_i}{2 \pi} \int_{\mathbb{R}^{3,1}} F_2 \wedge F_2,
\eeq
which combined with the result (\ref{FF}) for the CP-even coupling, yields the following expression for the gauge kinetic function:
\beq
f_{D7_i}= \frac{T_i}{2\pi}.
\eeq
On the other hand, taking one of the $F$'s to be the compact flux (denoted as $F^c_2$)
and the other to be with non-compact indices, we obtain (working in Einstein frame):
\beq
\C{L} \supset -\frac{1}{2 \pi l_s^4 } \left(\int_{D_i} \hat{D}_j \wedge F^c_2 \right) \int_{\mathbb{R}^{3,1}} Q_2^j \wedge F_2
= \frac{q_{ij}}{l_s^2} \int_{\mathbb{R}^{3,1}} Q_2^j \wedge F_2,
\label{QF}
\eeq
where $q_{ij}$ is the charge of the 2-form $Q_2^j$ under the $U(1)$ living on the divisor $D_i$.
Expanding the gauge flux $F^c_2$ in the basis of (1,1)-forms $\hat{D}_i$
as $F^c_2= f_c^i \hat{D}_i$, and defining the Calabi-Yau intersection numbers as:
\beq
k_{ijk}= \frac{1}{l_s^6} \int_{\mc{M}_6}\hat{D}_i\wedge\hat{D}_j\wedge\hat{D}_k,
\eeq
we end up with the following expression for the $U(1)$-charge $q_{ij}$:
\beq
q_{ij}=\frac{1}{2 \pi l_s^2} \int_{D_i} \hat{D}_j \wedge F^c_2
=\frac{f_c^k}{2 \pi l_s^6} \int_{\mc{M}_6} \hat{D}_i\wedge\hat{D}_j \wedge \hat{D}_k = \frac{f_c^k}{2 \pi} k_{ijk}.
\eeq
Therefore the gauge flux coefficients and the intersection numbers determine
which 2-forms couple to the Abelian gauge boson which lives on the divisor $D_i$.
Recalling that the 2-forms $Q_2^j(x)$ are 4D dual to the axions $b_j(x)$,
we realise that the K\"ahler moduli which get charged under the $U(1)$, are
those parameterising the volume of 4-cycles that intersect the 2-cycle supporting the gauge flux.
This is topologically equivalent to saying that the K\"{a}hler moduli which
get a $U(1)$-charge are a combination of 4-cycles
corresponding to the 4-cycle that is Poincar\'e dual to the 2-cycle on which the magnetic flux is turned on.
Due to the coupling (\ref{QF}), the $U(1)$ gauge boson becomes massive
by eating the axion (or an appropriate combination of axions) which is the imaginary part of
the charged K\"{a}hler modulus.

In order to see this mechanism in more detail, we need to include also the kinetic terms for the $Q_2^j$
which are expressed in terms of the corresponding field strength $H_3^j=d Q_2^j$. They can be derived from the 10D term $S \supset - \frac{1}{8\kappa_{10}^2} \int F_5 \wedge \star F_5$:
\beq
-\frac{2\pi}{4 l_s^6} \int_{\mathbb{R}^{3,1}\times\mc{M}_6} dC_4 \wedge * dC_4
= -\frac{\pi}{l_s^2} \left(\int_{\mc{M}_6} \hat{D}_j \wedge *\hat{D}_k \right)\int_{\mathbb{R}^{3,1}} \frac{1}{2} d Q_2^j \wedge * d Q_2^k,
\eeq
where:
\beq
\frac{1}{l_s^6} \int_{\mc{M}_6} \hat{D}_j \wedge *\hat{D}_k = \frac{(\mc{K}_0^{-1})_{jk}}{\vo} ,
\eeq
with $\vo$ the dimensionless Calabi-Yau volume.
The matrix $\mc{K}^{-1}_0$ is defined as the inverse of the metric obtained by taking the second derivatives of the tree-level
K\"{a}hler potential $K_0=-2\ln\vo$ with respect to the real part of the $T$-moduli. Thus we end up with:
\beq
-\pi \frac{(\mc{K}^{-1}_0)_{jk}}{\vo l_s^2 }\int_{\mathbb{R}^{3,1}} \frac{1}{2} d Q_2^j \wedge * d Q_2^k=
-\pi\frac{(\mc{K}^{-1}_0)_{jk}}{\vo l_s^2}\int_{\mathbb{R}^{3,1}} \frac{1}{12} H_{\mu \nu \rho}^j H^{k, \mu \nu \rho} d^4 x.
\label{HH}
\eeq
Our final Lagrangian is then given by the standard Maxwell action (\ref{FF}) plus the
term (\ref{QF}) describing the coupling of the 2-form $Q_2^j$ to the Abelian gauge boson
and the 2-form kinetic term (\ref{HH}). Before showing how the gauge boson becomes massive,
let us redefine the 2-form so that it gets a canonical mass dimension 1:
\beq
Z_2^j \equiv M_P Q_2^j\,\,\,\,\,\,\,\,\Leftrightarrow\,\,\,\,\,\,\,\,G_3^j \equiv M_P H_3^j.
\eeq
Using $l_s^{-2}=M_s^2=M_P^2/(4\pi\vo)$, the final Lagrangian takes the form:
\beq
\mathcal{L}=-\frac{(\mc{K}^{-1}_0)_{jk}}{48\vo^2} G_{\mu \nu \rho}^j G^{k, \mu \nu \rho}
-\frac{1}{4 g_i^2} F_{\mu\nu}F^{\mu\nu} +q_{ij} \frac{M_P}{4\pi \vo} Z_2^j \wedge F_2.
\eeq
The 2-form kinetic terms can be canonically normalised by defining (suppressing the space-time indices):
\beq
G^i=2\vo\, C^i_j \mc{H}^j,\,\,\,\,\Leftrightarrow\,\,\,\,Z_2^i=2\vo\, C^i_j \mc{D}_2^j,
\eeq
where the columns of the matrix $C^i_j$ are given by the eigenvectors of $K^{-1}_0$ normalised as:
\beq
(\mc{K}_0^{-1})_{ij} C^j_a =  C_{ia}\lambda_a,\text{ \ \ \ with \ \ \ }(C^t)_a^i C_{ib}= \lambda_a^{-1} \delta_{ab}.
\eeq
Note that
\beq
(C^t)_b^i (\mc{K}_0^{-1})_{ij} C^j_a = \delta_{ab} \rightarrow (C^t)_b^i (\mc{K}_0^{-1})_{ij} C^j_a (C^t)^a_k = (C^t)^b_k \rightarrow C^j_a (C^t)^a_k = (\mc{K}_0)_{ik}.
\label{InverseId}\eeq
Canonically normalising also the $U(1)$ field strength as $F= g_i \mc{F}^i$, and using the expressions (\ref{FF}) and (\ref{U1charge})
for the coupling constant $g_i$ and the $U(1)$-charge $q_{ij}$ respectively, we end up with
(defining the dimensionless flux coefficients $f^i$ as $f^i\equiv l_s^2 f_c^i/(2 \pi)$):
\beq
\mathcal{L}=-\frac{1}{12} \mc{H}_{\mu \nu \rho}^j \mc{H}_j^{\mu \nu \rho}
-\frac{1}{4} \mc{F}_{\mu\nu}^i\mc{F}^{i\,\mu\nu}
+ M_{ij}\mc{D}_2^j \wedge \mc{F}_2^i.
\label{Lfin}
\eeq
where:
\beq
M_{ij}\equiv  \left(g_i f^k k_{ipk} C^p_j \right) \frac{M_P}{2\pi}
=\left( g_i \,q_{ip} C^p_j \right) \frac{ M_P}{2\pi}.
\eeq
with no sum over $i$ since this index simply denotes the 4-cycle $D_i$ wrapped by the $D7$-brane.
Hence we realise that in general $F_2$ couples to a particular combination of all the 2-forms, and
not just to a single 2-form, due to the canonical normalisation which typically
introduces a mixing among all the 2-forms.

Let us see why on dualising $\mc{D}_2$ to the corresponding axion $a$,
the Lagrangian (\ref{Lfin}) generates an explicit mass term $m_{\gamma^\prime}^2 \mc{A}_\mu \mc{A}^\mu$ for the $U(1)$ gauge boson.
The dual axion $a$ can be introduced as a Lagrange multiplier for the arbitrary field $\mc{H}_{\mu\nu\rho}$
by imposing the constraint $d^* \mc{H}=0$ which is locally equivalent to $d\mc{D}_2=\mc{H}$:
\beq
\mathcal{L}=-\frac{1}{12} \mc{H}_{\mu \nu \rho}^j \mc{H}_j^{\mu \nu \rho}
-\frac{1}{4} \mc{F}_{\mu\nu}\mc{F}^{\mu\nu}
- \frac{M_{ij}}{6}\ \epsilon^{\mu\nu\rho\sigma} \mc{H}^j_{\mu\nu\rho}\ A_{\sigma}
-\frac{M_{ij}}{6}\,a \epsilon^{\mu\nu\rho\sigma} \partial_\mu \mc{H}^j_{\nu\rho\sigma}.
\label{Ldual}
\eeq
We can now obtain a quadratic Lagrangian for $\mc{H}$
by integrating by parts the last term in (\ref{Ldual}).
Then the equations of motion for $\mc{H}$ give:
\beq
\mc{H}_j^{\mu\nu\rho}= - {M_{ij}}\ \epsilon^{\mu\nu\rho\sigma}
\left(A_\sigma+\partial_\sigma a\right),
\eeq
which inserted back into (\ref{Ldual}) yields:
\beq
\mc{L}= -\frac{1}{4}\mc{F}_{\mu\nu}\mc{F}^{\mu\nu} -
\frac{m_{\gamma^\prime}^2}{2} \left(A_\mu+\partial_\mu a\right) \left(A^\mu+\partial^\mu a\right),\text{ \ \ \ with \ \ \ }m_{\gamma^\prime}^2=\sum_j M_{ij}^2.
\eeq
The field $\mc{A}_{\mu}\equiv A_\mu+\partial_\mu a$ clearly represents a massive $U(1)$ gauge boson.
Thus the axion $a$ is eaten up by the gauge boson without the need of any Higgs-like field
in a stringy realisation of the standard St\"uckelberg mechanism.
Note that the above can be simplified using \ref{InverseId} to
\begin{align}
(M^2)_{ab} =& M_{aj} M^t_{jb} = \frac{ M_P^2}{2\pi\sqrt{\tau_a\tau_b} } q_{ap} C^p_j (C^t)^j_r q_{br} \nonumber \\
=&  \frac{M_P^2}{2\pi\sqrt{\tau_a\tau_b} } q_{ap} q_{br} (\mc{K}_0)_{pr}.
\end{align}

If we include also the contributions from four-cycles odd under the orientifold $\hat{D}^c_-$, defining
\begin{align}
 r_{a c} \equiv \int_{D_a} \hat{D}^c_-,
\end{align}
the final total result is
\begin{align}
\label{StmassTotal}
m_{ab}^2=& g_a g_b\,  \frac{M_{P}^2}{4\pi^2} \bigg[ \V^{-2} r_{a c}  (\C{K}_0^{-1})^{cd} r_{b d} + q_{a \alpha} (\C{K}_0)_{\alpha \beta}  q_{b \beta}\bigg].
\end{align}
This expression is equivalent to the one presented in \cite{Goodsell:2009xc} noting the different metrics used and the different definition of $M_s$.


\subsection{FI-terms}

In the previous section we have seen how $U(1)$ gauge bosons can acquire a mass by turning on
an internal magnetic flux in the world-volume of a $D7$-brane wrapping a divisor $D_i$
with corresponding K\"{a}hler modulus $T_i$. This guarantees
that also a moduli-dependent 4D Fayet-Iliopoulos term gets generated \cite{Jockers:2005zy,Haack:2006cy,Dine:1987xk}.
In fact, denoting as $T_{U(1)}$ the charged K\"{a}hler modulus which is in general a
combination of all the basis divisors corresponding to the 4-cycle Poincar\'e dual to the 2-cycle supporting
the magnetic flux, the axion $a=\text{Im}(T_{U(1)})$ gets eaten up by the $U(1)$ gauge boson
via the St\"{u}ckelberg mechanism, but $\tau=\text{Re}(T_{U(1)})$ is a light modulus
that has to be taken into account in the effective field theory and
gives rise to a moduli-dependent Fayet-Iliopoulos term.

This can be seen to arise from the low-energy reduction of the DBI action (\ref{DBIS}):
\beq
S_{DBI}=-\mu_7 e^{-\phi}\Gamma(y)\int_{\mathbb{R}^{3,1}}d^4 x
\left(1+ \frac{l_s^4}{16 \pi^2}F_{\mu\nu}(x)F^{\mu\nu}(x)+...\right), \label{DBIred}
\eeq
where:
\beq
\Gamma(y)=\int_{D_i}d^4 y\sqrt{-\textrm{det}\left[\varphi^*(g_{CY})\right]}
\left(1+ \frac{l_s^4}{16 \pi^2}F_{mn}(y)F^{mn}(y)+...\right). \notag
\eeq
From the BPS calibration condition for a $D7$-brane we find that:
\beq
\Gamma(y)=\frac{1}{2}\int_{D_i}\left(J\wedge J-\frac{l_s^4}{4\pi^2} F_2^c\wedge F_2^c\right)
+\frac{\left(\int_{D_i}J\wedge \frac{l_s^2}{2\pi}F_2^c\right)^2}
{\int_{D_i}\left(J\wedge J-\frac{l_s^4}{4\pi^2} F_2^c\wedge F_2^c\right)}.
\eeq
When in (\ref{DBIred}) $\Gamma(y)$ multiplies the first term in parenthesis, after performing the appropriate
Weyl rescaling to 4D Einstein frame\footnote{We recall that the 10D metric in string frame is related
to the 10D metric in Einstein frame via $g_{MN}^{(s)}=e^{\phi/2}g_{MN}^{(E)}$.}, we obtain two contributions to the 4D scalar potential:
the $D7$-brane tension $T_{D7}$ and a moduli dependent
Fayet-Iliopoulos term $\xi_i$:
\beq
T_{D7}= g_i^{-2} 4\pi^2 e^{2\phi} M_s^4, \text{ \ \ and \ \ }V_D=\frac{g_i^2}{2}\xi_i^2,
\text{ \ \ with \ \ }\frac{\xi_i}{M_P^2}=\frac{1}{4\pi\vo}\int_{D_i}\left(J\wedge \frac{l_s^2}{2\pi}F^c_2\right).
\label{fi}
\eeq
The $D7$-brane tension gives no net contribution to the scalar potential since it will be compensated by
by other extended objects due to tadpole cancellation.

On the other hand, considering in (\ref{DBIred}) $\Gamma(y)$ multiplied by the second term in parenthesis,
we realise that in the presence of a non-vanishing world-volume flux,
the expression (\ref{FF}) for the gauge coupling constant $g_i$ gets modified to:
\beq
\frac{2\pi}{g_i^2} = \text{Re}(T_i) - h_i (F_2^c)  \text{Re}(S),
\label{g-2}
\eeq
where $\text{Re}(S)=e^{-\phi}$ and the flux-dependent factor is given by $h_i (F_2^c)= \frac{f^k f^j k_{ijk}}{2} = \frac{f^j q_{ij}}{2}$
where $q_{ij}$ are the flux-dependent $U(1)$ charges of the K\"{a}hler moduli (\ref{U1charge}).

The Fayet-Iliopoulos term in (\ref{fi}) can be rewritten as:
\beq
\frac{\xi_i}{M_P^2}=\frac{1}{4\pi\vo}\int_{D_i}\left(J\wedge \frac{l_s^2}{2\pi}F^c_2\right)
=\frac{1}{4\pi\vo}t^j f^k k_{ijk}= \frac{q_{ij}}{4\pi}\frac{t^j}{\vo}=-\frac{q_{ij}}{4\pi}\frac{\partial K}{\partial \tau_j}.
\label{FI}
\eeq
Including also the presence of unnormalised charged matter fields $\varphi_j$ (open string states)
with corresponding $U(1)$ charges given by $c_{ij}$, the resulting $D$-term potential looks like
(considering the dilaton fixed at its VEV: $e^{\phi}=g_s$):
\beq
V_D = \frac{g_i^2}{2} \left( \sum_j c_{ij} \varphi_j \frac{\partial K}{\partial \varphi_j} -\xi_i\right)^2
=\frac{\pi}{\left(\tau_i-f^j q_{ij}/(2g_s)\right)} \left( \sum_j c_{ij} \varphi_j \frac{\partial K}{\partial \varphi_j}
+\frac{q_{ij}}{4\pi}\frac{\partial K}{\partial \tau_j}\right)^2.
\label{VD}
\eeq

\providecommand{\href}[2]{#2}\begingroup\raggedright\endgroup

\end{document}